# Linear and nonlinear benchmark of gyrokinetic simulation of energetic particle driven toroidal Alfven eigenmodes in ITPA TAE benchmark case

BY YOUJUN HU[1], YANG CHEN[2], LEI YE[1], ZHIYONG QIU[1], YOUWEN SUN[1]

1. Institute of Plasma Physics, Hefei Institutes of Physical Science, Chinese Academy of Science, Hefei 230031, China
2. Department of Physics, University of Colorado, Boulder, CO 80309, USA

**Abstract**

A new gyrokinetic code, `TEK`, was benchmarked in simulating energetic particle (EP) driven toroidal Alfven eigenmodes (TAEs) in the simple tokamak configuration chosen by the ITPA-EP group for code benchmarking purpose. Linear benchmark has been well established by other codes, whereas nonlinear benchmark for this case is lacking. This paper presents, besides the linear benchmark, nonlinear results for both single-$n$ and multiple-$n$ simulations ($n$ is the toroidal mode number). The nonlinear results are in good agreement with an analytical theory on zonal field beat-driven by Alfven eigenmodes, partially verifying correctness of the nonlinear simulations. The saturation level and the resulting EP transport are examined. This provides data for future inter-code nonlinear benchmarking.

In `TEK`, all species (electrons, thermal ions, EPs) are treated on the same footing using the gyrokinetic model (with electrons in the zero Larmor radius limit). The electromagnetic cancellation problem is mitigated by using the mixed-variable pullback method. Numerical details related to electromagnetic gyrokinetic simulation are discussed.

## 1 Introduction

Deuterium–tritium fusion born $\alpha$ particles and auxiliary heating generated energetic ions in tokamaks are required to have good confinement so that they provide efficient plasma heating and do not damage the machine wall. Meanwhile, these energetic particles (EPs) can excite a kind of electromagnetic perturbations called toroidal Alfven eigenmodes (TAEs)[1], which may grow to significant amplitude to impact the dynamics of the EPs, thus impacting their phase-space distribution, heating efficiency, radial transport, and loss to the wall[2].

To benchmark the reliability of various codes in simulating linear dynamics of TAEs, the ITPA-EP group chose a simple circular tokamak and focus on the $n=6$ TAE, where $n$ is the toroidal mode number. This benchmark practice has been well documented in Ref. [3]. While linear simulations are useful (e.g., in determining mode structures), they can not give amplitude of the modes, and thus are unable to answer whether the modes will be important in affecting the EP dynamics. For this, we need nonlinear simulation. However nonlinear benchmark for this case is lacking due to difficulties in benchmarking nonlinear simulations. These difficulties appear for some reasons (e.g. nonlinear simulations are more sensitive to details of the physical model). In this work, as a first step toward inter-code benchmark, we compare nonlinear simulations with Liu Chen's analytical theory on zonal field beat driven by Alfven eigenmodes[4]. The comparison show good agreement between simulation and theory, giving us confidence in the correctness of the nonlinear simulations. The saturation level and the resulting EP transport are then examined, which indicates the TAE induces negligibly small EP transport in this particular case. These provide data for future inter-code nonlinear benchmarking.

The simulations presented in this paper were performed by using a new $\delta f$ gyrokinetic PIC code, `TEK`, in which the electromagnetic cancellation problem is mitigated by using the mixed-variable pullback method [5, 6, 7]. Numerical details related to electromagnetic gyrokinetic simulation are discussed (see Appendix A), which may be useful for those who are interested in developing gyrokinetic codes. `TEK` has also been benchmarked with the `GENE` code[8] for the ITG-KBM transition in the DIII-D cyclone base (see Appendix B), which demonstrates its capability of handling other electromagnetic modes (besides shear Alfven waves discussed here) using the same algorithm.

# 2 Equilibrium magnetic configuration and species profiles used in ITPA-EP TAE benchmarking

The simple configuration chosen by the ITPA-EP group for code benchmarking purpose is a circular tokamak with the magnetic surfaces given by

$$\begin{aligned} R(r,\theta) &= R_0 + r\cos\theta, \\ Z(r,\theta) &= r\sin\theta, \end{aligned} \tag{1}$$

where $(R,\phi,Z)$ are the right-handed cylindrical coordinates, $(r,\theta)$ are the minor radius and poloidal angle, with $\theta$ in the range $[-\pi:\pi)$ and $r$ in the range $[0:a)$, where $a=1m$ is the minor radius of the boundary magnetic surface. $R_0=10m$ is the major radius of the magnetic axis. Other quantities required to fully specify the configuration are given in Table 1.

| $q(r)$ | $g(r)\equiv B_\phi R$ |
|---|---|
| $1.71+0.16(r/a)^2$ | $30Tm$ |

**Table 1.** Here $B_\phi$ is the toroidal component of the magnetic field. The toroidal field function $g(r)$ is assumed to be a radial constant. $q(r)$ is the safety factor defined by $q=(2\pi)^{-1}\int_0^{2\pi}\mathbf{B}\cdot\nabla\phi/(\mathbf{B}\cdot\nabla\theta)d\theta$.

Using the above information, the poloidal magnetic flux can be calculated as

$$\Psi_p(r)=\Psi_p(0)+\int_0^r \frac{1}{q(r)}\frac{2\pi g r}{\sqrt{R_0^2-r^2}}dr,$$

and the toroidal magnetic flux

$$\Psi_t(r)=2\pi g\left(\sqrt{R_0^2}-\sqrt{R_0^2-r^2}\right).$$

Then the magnetic field is written as

$$\begin{aligned}\mathbf{B}_0 &= \frac{1}{2\pi}\nabla\Psi_p\times\nabla\phi+g\nabla\phi \\ &= \nabla r\times\nabla\phi\frac{1}{q(r)}\frac{gr}{\sqrt{R_0^2-r^2}}+g\nabla\phi\end{aligned}$$

In this setup, the toroidal plasma current $I_\phi < 0$ and toroidal magnetic field $B_\phi > 0$. Specifically, the toroidal plasma current $I_\phi$ within the boundary surface $(r=1m)$ is $-810$kA, and $B_\phi$ at the magnetic axis is $+3$Tesla.

Define

$$\rho_t=\sqrt{\frac{\Psi_t(r)}{\Psi_t(a)}}=\sqrt{\frac{R_0-\sqrt{R_0^2-r^2}}{R_0-\sqrt{R_0^2-a^2}}}, \tag{2}$$

which is the square root of the normalized toroidal magnetic flux. We will use $\rho_t$ as the radial coordinate (rather than $r/a$) when presenting the simulation results. In the limit of $r/a\to 0$, then $\rho_t = r/a$. In the general case, the expression of $r$ in terms of $\rho_t$ can be solved form Eq. (1), yielding $r = \sqrt{R_0^2-\left[R_0-\rho_t^2\left(R_0-\sqrt{R_0^2-a^2}\right)\right]^2}$.

The poloidal flux data ($\Psi=\Psi_p/2\pi$, i.e., poloidal flux per rad) are written as a numerical G-eqdsk file, which is then read in by `TEK`. This enables us to avoid using any analytical formulas for the magnetic field. I.e., we treat this analytical configuration as a general numerical one.

The plasma consists of 3 species: electrons, thermal ions (protons), and EPs (Deuterons). All 3 species are of Maxwellian in the velocity space. The temperature of the 3 species is radially uniform. The thermal ion density is also radially uniform ($n_i = 2\times 10^{19} m^{-3}$). The EP density is radially nonuniform with a profile given by an analytical formula (see Table 2). The electron density is set to satisfy the charge neutrality: $n_e = n_i - n_{\rm EP}$. `TEK` simulations indicate the TAE frequency is sensitive to the thermal plasma density profile: a slight decrease in $n_e$ ( e.g., 1%) will result a large decrease in the TAE frequency (e.g., 10%). Examples will be given in Sec. 3.

To avoid ambiguity, the detailed parameters for the 3 species are listed in Table 2.

| | Thermal ion: Proton | EP: Deuteron | Electron |
|---|---|---|---|
| Mass (kg) | $1.67355\times 10^{-27}$ | $3.34358\times 10^{-27}$ | $9.10938\times 10^{-31}$ |
| Charge (e) | +1 | +1 | -1 |
| Temperature (keV) | 1 | 400 | 1 |
| Number density ($m^{-3}$) | $2\times 10^{19}$ | $n_0 c_3 \exp\left[-\frac{c_2}{c_1}\tanh\left(\frac{\rho_t-c_0}{c_2}\right)\right]$ | $n_i - n_{\rm EP}$ |

**Table 2.** Parameters for the 3 plasma species. The parameters appearing in the EP density profile are given by $n_0 = 1.44131\times 10^{17}$, $c_0 = 0.49123$, $c_1 = 0.298228$, $c_2 = 0.198739$, and $c_3 = 0.521298$. All the parameters are the same as in Ref. [3].

For the linear simulations, we scan the EP temperature (from 100keV to 800keV) to examine the dependence of TAE frequency and growth rate on it. For the nonlinear simulations, we consider only the case of the EP temperature being 400keV (the baseline case). All the 3 species (electrons, thermal ions, EPs) are treated on the same footing using the gyrokinetic model (with electrons in the zero Larmor radius limit). The finite Larmor radius (FLR) effect is always kept for species other than electrons (namely thermal ions and EPs) in this paper.

## 3 Linear simulations

In this section, we focus on linear simulations of $n=6$ TAE, and examine its mode structure, frequency, and growth rate.

First we consider the base-line case (where the EP temperature is 400keV). Figure 1 upper panel plots a snapshot of $\delta\Phi$ and $\delta A_{\parallel}$ in a poloidal plane, where $\delta\Phi$ is the scalar potential perturbation, and $\delta A_{\parallel}$ is the parallel vector potential perturbation. In this paper, $\delta\Phi$ is normalized by $T_u/e$, and $\delta A_{\parallel}$ is normalized by $T_u/(ev_u)$, where $e$ is the elementary charge, $T_u=1$keV, and $v_u=1.5237\times10^7 m/s$.

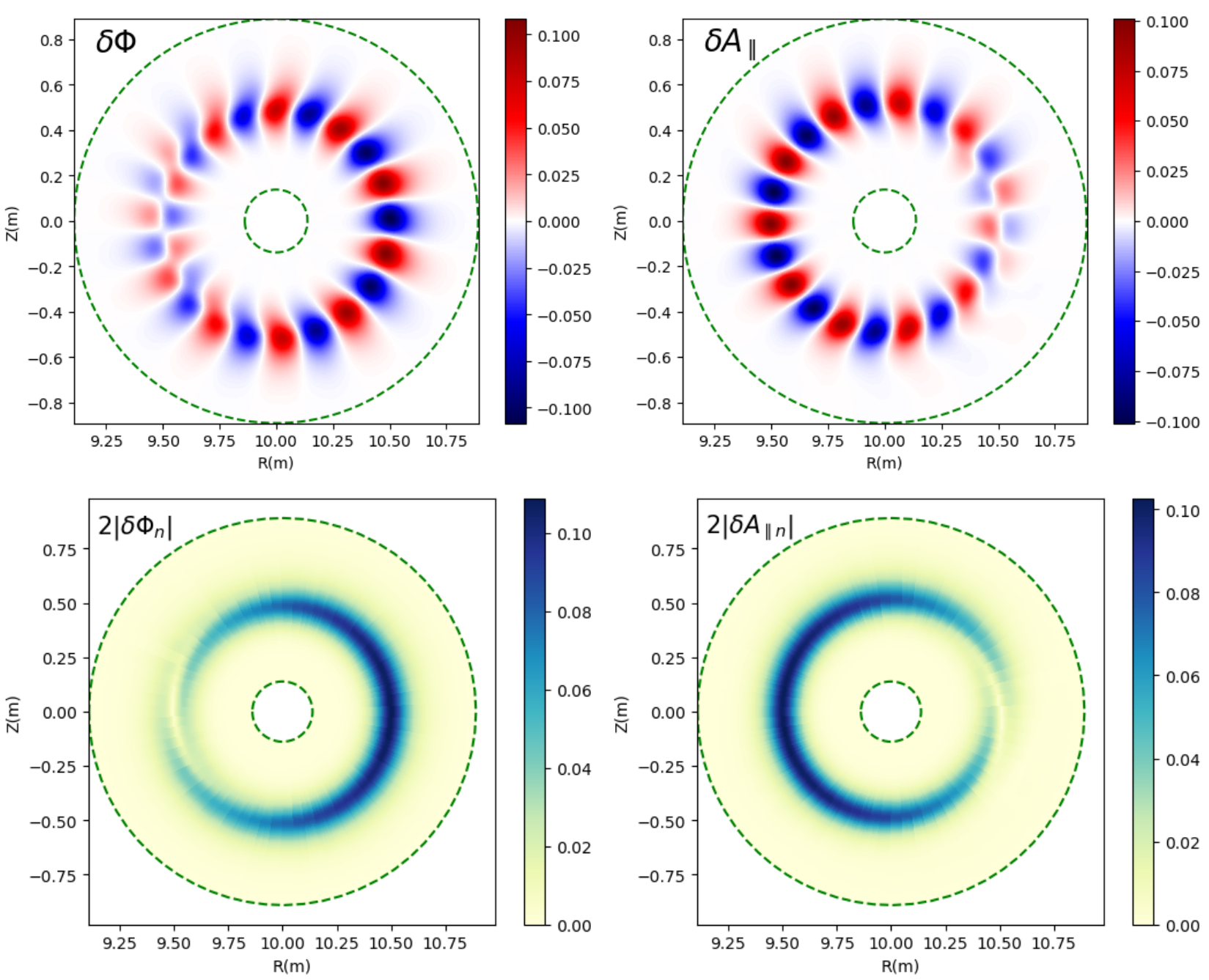

**Figure 1.** Upper panel: Snapshot of $\delta\Phi$ and $\delta A_{\parallel}$ in $\phi=0$ poloidal plane. Lower panel: Amplitude of the $n=6$ toroidal harmonic. The dashed lines indicate the radial computational boundaries. Here $\delta\Phi_n$ is defined by the Fourier expansion $\delta\Phi(\phi)=\sum_{n=-\infty}^{n=+\infty}\delta\Phi_n\exp(in\phi)$, and similar definition for $\delta A_{\parallel n}$. One physical mode is composed of two harmonics of $n=+6$ and $n=-6$ in basis functions $\exp(in\phi)$. Summing these two harmonics gives the physical $n=6$ mode, whose amplitude is twice $|\delta\Phi_n|$.

Figure 1 lower panel plots the amplitudes of the $n=6$ toroidal harmonics of $\delta\Phi$ and $\delta A_\parallel$. The results show that the amplitudes are not poloidally symmetric. This poloidal asymmetry is a typical feature of TAEs. A TAE consists of two adjacent poloidal harmonics: say $m$ and $m+1$, where $m$ is poloidal mode number. The superposition of the two harmonics implies that: if they interferes each other constructively at one poloidal angle, then after $\pi$ poloidal angle variation they will interfere destructively. The results in Fig. 1 indicate that for $\delta\Phi$ the region of interfering constructively is near the low-field-side (ballooning structure), and for $\delta A_\parallel$ the region is near the high-field-side (anti-ballooning structure).

To see the poloidal ballooning structure more clearly, Fig. 2 plots a 1D snapshot of $\delta\Phi$ and $\delta A_\parallel$ along the poloidal direction for a fixed radial location ($\rho_t=0.5$). Also plotted are the amplitudes, which form the envelope of the instantaneous values. (Figure 2 is just a 1D cross section of Fig. 1 taken at $\rho_t=0.5$.) The shape of the envelope remains the same (magnitude grows) during the TAE linear growth, indicating the poloidal group velocity is zero.

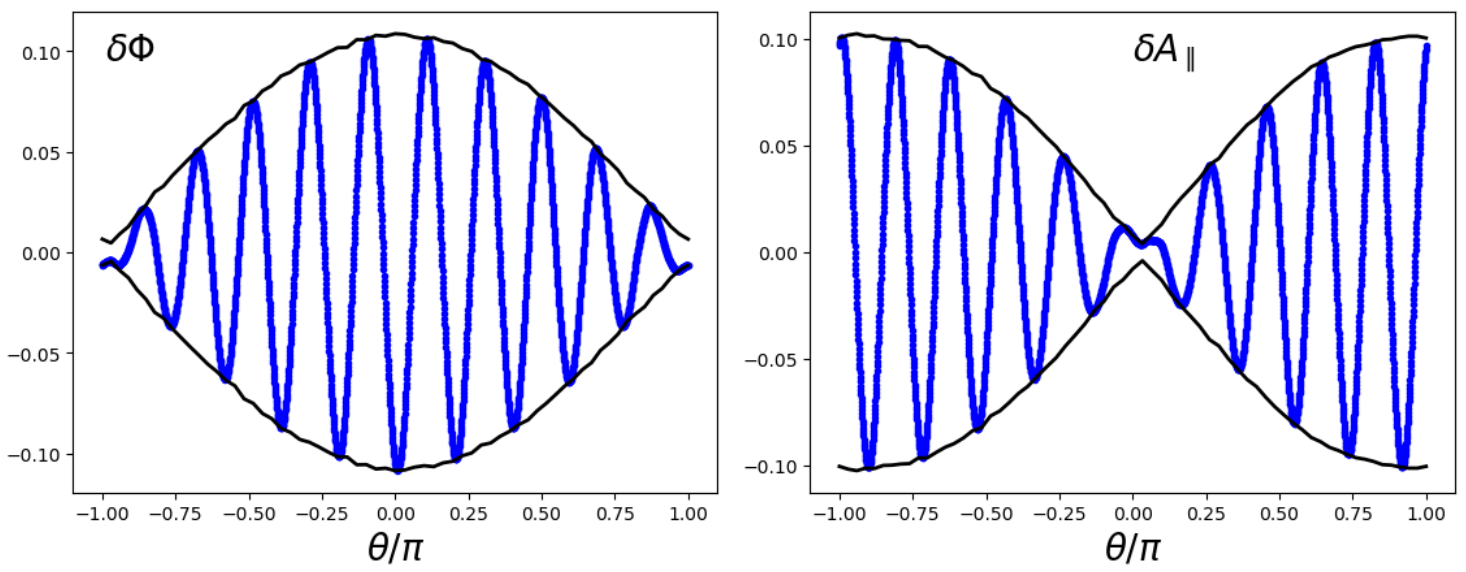

**Figure 2.** Snapshot of $\delta\Phi$ and $\delta A_\parallel$ along the poloidal direction for a fixed radial location ($\rho_t=0.5$). The black lines are the amplitudes, $2|\delta\Phi_n|$ and $2|\delta A_{\parallel n}|$. Here $\theta=0$ corresponds to the low-field-side, and $\theta=\pm\pi$ the high-field side.

The poloidal phase velocity is in the anti-clockwise direction (viewed along $\nabla\phi$). This is identical to the ion diamagnetic drift direction (i.e., $\mathbf{B}\times\nabla p$). The toroidal phase velocity direction is along $-\nabla\phi$, which is the toroidal plasma current direction. These results agree with the general rules about the phase velcoity of TAEs excited by fast ions [9]: poloidally along the ion diamagnetic drift and toroidally in the plasma current direction (co-current). If one wants to express the phase velocity direction in terms of a spatiotemporal phase, e.g. $\omega t - m'\theta + n\phi$ with $\omega>0$, then the above direction corresponds to $m'>0$ and $n>0$.

Decomposing the 2D snapshot into poloidal harmonics at each radial location, we get the radial structure of various poloidal harmonics. Figure 3 plots the amplitude of the $m=9,10,11,12$ harmonics. All other harmonics that are not plotted here are negligibly small. The results indicate that two adjacent harmonics, namely $m=10$ and $m=11$, are dominant over other harmonics.

This is consistent with what we expect from a TAE mode. Figure 3 agrees with simulations performed by `TRIMEG-GKX` code: Figure 5 in Ref. [10].

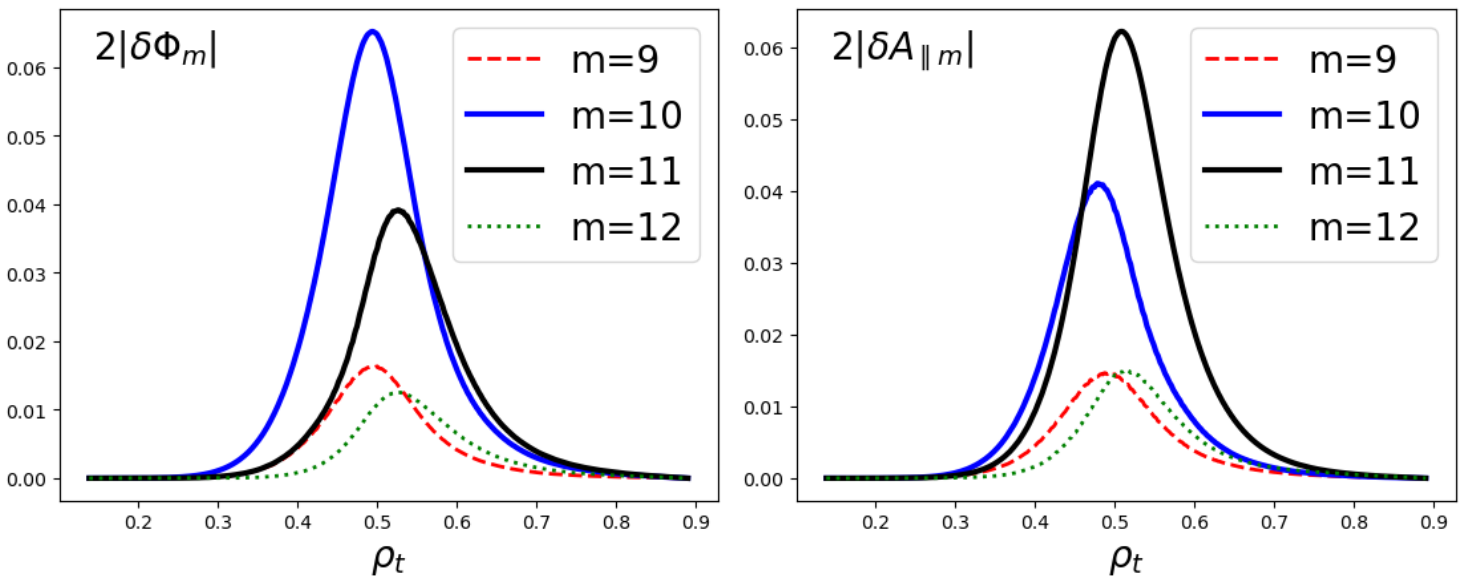

**Figure 3.** Radial profiles of amplitudes of $m = 9, 10, 11, 12$ poloidal harmonics. Left panel is for $\delta\Phi$, and the right panel is for $\delta A_{\parallel}$. Here $\delta\Phi_m$ is defined by $\delta\Phi(\theta) = \sum_{m=-\infty}^{m=+\infty} \delta\Phi_m \exp(im\theta)$, and similar definition for $\delta A_{\parallel m}$.

Meanwhile, it is important to note that the relative amplitude shape among poloidal harmonics depends on the definition of the poloidal angle. The poloidal angle $\theta$ appears in Eq. (1) is the usual geometric poloidal angle, and is what is used in decomposing the mode into poloidal harmonics in Fig. 3. Another often used poloidal angle is the straight-field-line poloidal angle $\theta_f$, which for the simple configuration discussed here is related to $\theta$ by[11]

$$\theta_f = 2\arctan\left(\frac{(R_0 - r)}{\sqrt{R_0^2 - r^2}}\tan\left(\frac{\theta}{2}\right)\right). \tag{3}$$

Use this angle to do the poloidal harmonics decomposition, then the amplitude shape will be different from that in Fig. 3. The results are shown in Fig. 4. In this case, the two dominate harmonics are almost of the same amplitude, which is more close to the ideal picture of TAEs. So we need to make sure we are using the same poloidal angle when we compare simulations between different codes. Otherwise, a coordinate transform is needed before comparing.

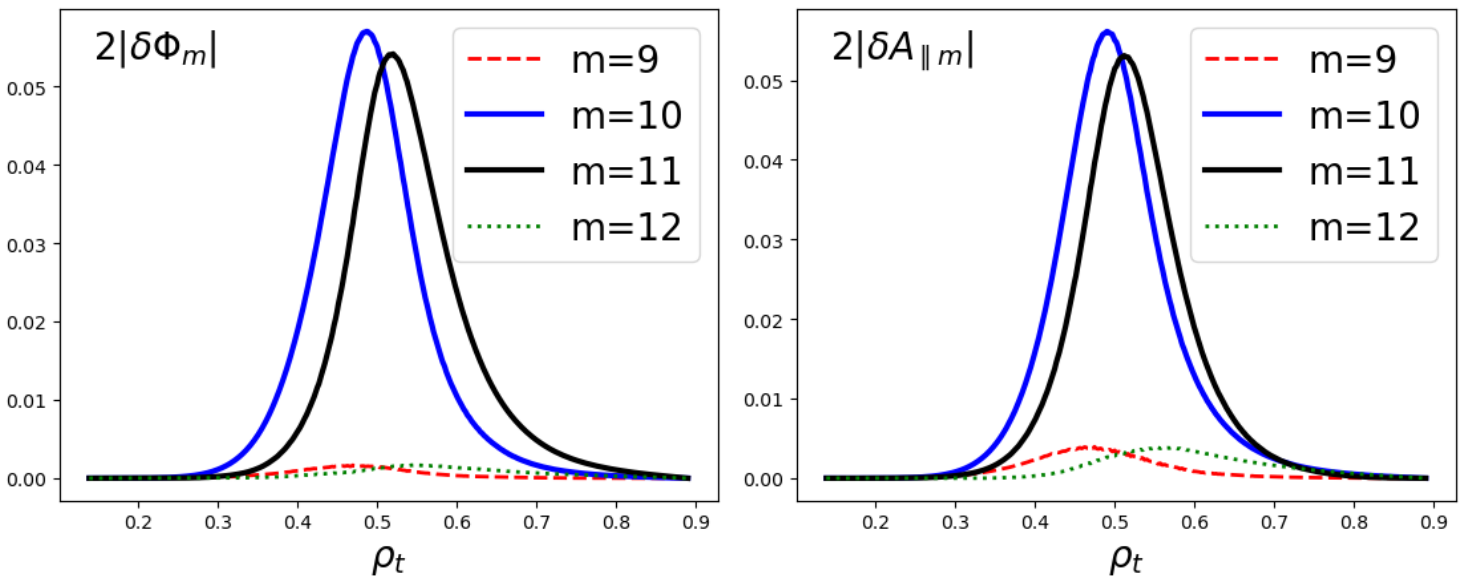

**Figure 4.** Radial profiles of amplitudes of poloidal harmonics using the straight-field-line $\theta_f$. Left panel is for $\delta\Phi$, and the right panel is for $\delta A_{\parallel}$.

Comparing TEK simulations with simulations presented in Könies' benchmarking paper [3], we found some simulations (e.g., LIGKA[12], MEGA[13], CAS3D[14], CKA, AE3D, CASTOR) agree with Fig. 4, while other simulations (e.g. ORB5[15], HMGC[16], EUTERPE[17], GYGLES[18]) agree with Fig. 3. So by inference we know the former codes were using the straight-field-line poloidal angle, while the latter were using the usual geometric poloidal angle. In the rest of this paper, all poloidal harmonics decompositions by default use the straight-field-line poloidal angle $\theta_f$.

Figure 5 plots the frequency spectrum of $\delta\Phi$ and $\delta A_{\|}$ on the low field side (i.e., $\theta = 0$). The results show that $\delta\Phi$ spectrum is well-localized in the TAE gap, both in the radial range and frequency range. The hot spot appears around $\rho_t \approx 0.5$ and $\omega \approx 425\text{kHz}\cdot\text{rad}$. Meanwhile we note that $\delta A_{\|}$ spectrum shifts slightly away form $\rho_t \approx 0.5$, with the central angular frequency being still $425\text{kHz}\cdot\text{rad}$.

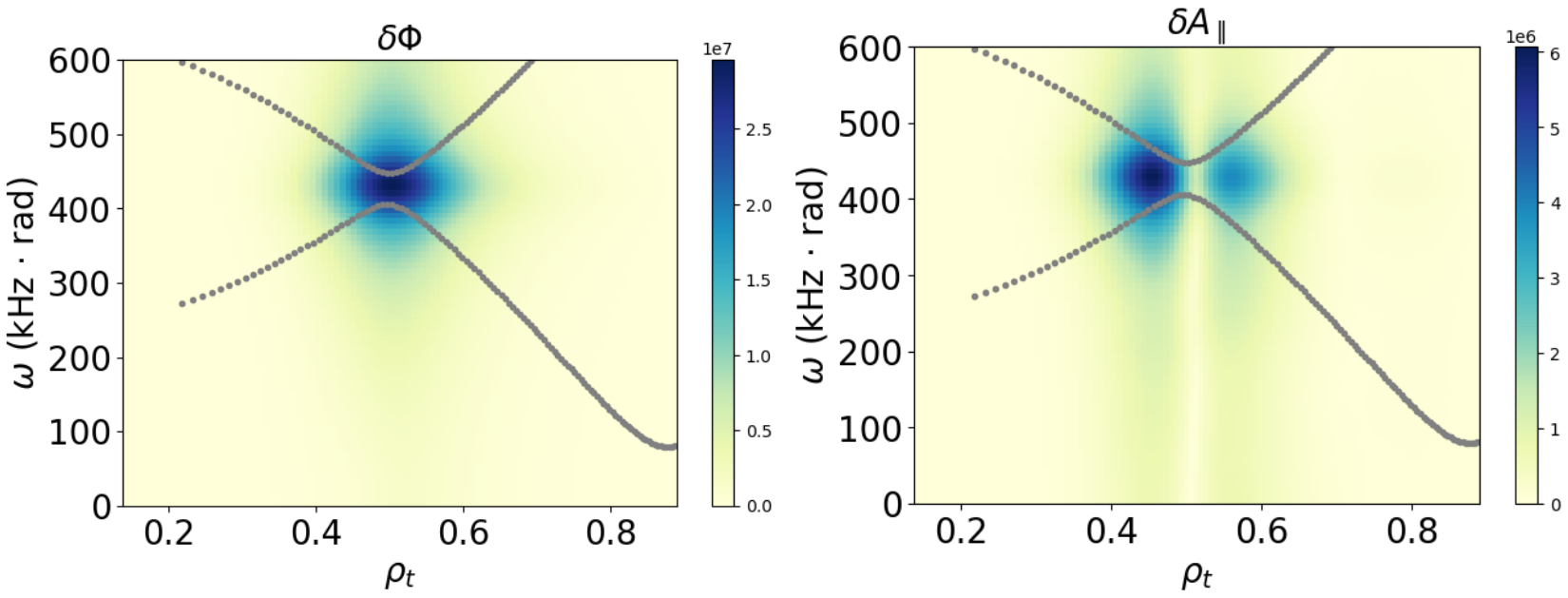

**Figure 5.** Angular frequency spectrum of $\delta\Phi(\rho_t, \theta, \phi)$ (left panel) and $\delta A_{\|}(\rho_t, \theta, \phi)$ (right panel) at $(\theta = 0, \phi = 0)$. The gray dots are $n = 6$ MHD toroidal Alfven continua calculated by using the GTAW code[19] in the slow-sound approximation.

To show more details of the TAE gap, Fig. 6 plots the $n = 6$ toroidal Alfven continuum along with two cylindrical Alfven continua ($m = 10$ and $m = 11$). This indicates the gap is formed by the coupling between $m = 10$ and $m = 11$ harmonics, as is confirmed by the crossing of the two cylindrical continua at the gap center. Considering that the dominate poloidal harmonics identified above are $m = 10$ and $m = 11$, and the radial location and frequency is well localized in the TAE gap, we can conclude that the mode found in the simulation is indeed a TAE.

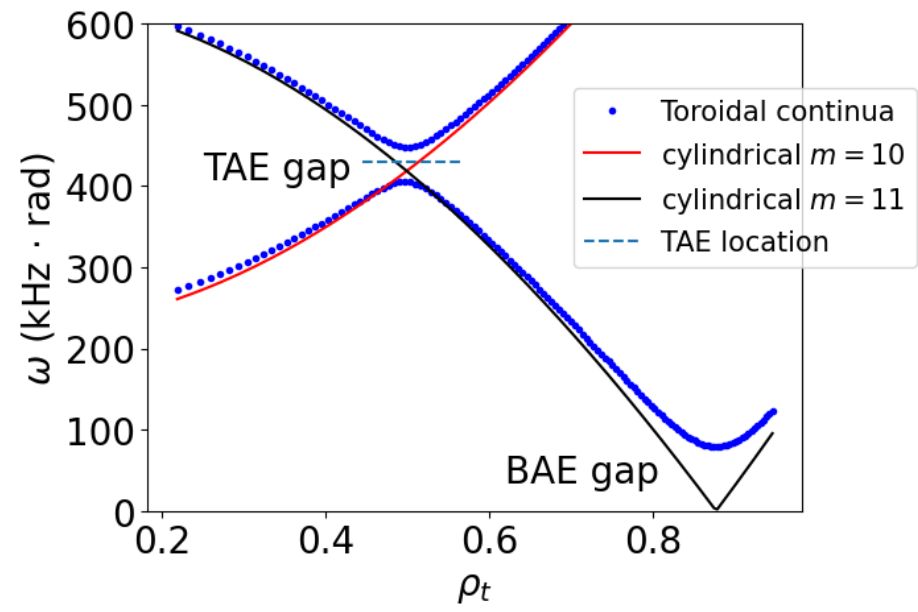

**Figure 6.** $n = 6$ toroidal and cylindrical Alfven continua . The dashed line indicates the angular frequency and half-height radial width of the mode found in the simulation. The intersect of $m = 11$ cylindrical continua with the horizontal axis is the resonant surface for $m = 11$ harmonic. The $m = 10$ harmonic does not have a resonant surface in the plasma.

As is mentioned above, TEK simulations indicate the TAE frequency is sensitive to the electron density profile. If we ignore the charge neutrality, and use a constant electron density profile, $n_e = n_i = 2 \times 10^{19} m^{-3}$, then the frequency spectrum is given by Fig. 7, which shows that the central $\omega$ is $376\text{kHz} \cdot \text{rad}$, about 12% reduction relative to the value $\omega = 425\text{kHz} \cdot \text{rad}$ given by Fig. 5. The frequency is now below the TAE gap tip, seemingly suggesting it is a continuum mode. But the mode structure (not shown) is similar to that shown above, indicating it is still a TAE.

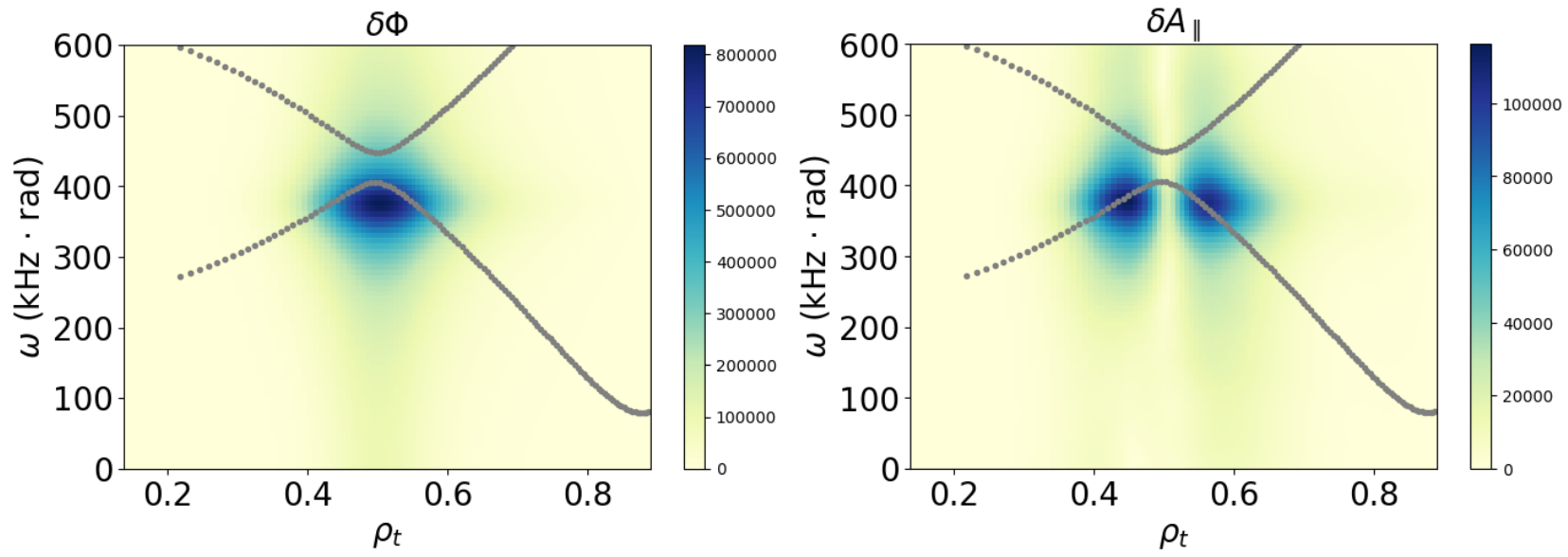

**Figure 7.** The same as Fig. 5 except that the electron density is changed to be uniform with $n_e = 2 \times 10^{19} m^{-3}$.

Next, we scan the EP temperature to examine the dependence of the TAE frequency and growth rate on it (we go back to using the electron density satisfying the charge neutrality). The results are plotted in Fig. 8, and are compared with two gyrokinetic codes: ORB5[15] and EUTERPE[17]. The results show that TEK results are in reasonable agreement with them. The FLR effect was included in all the 3 gyrokinetic codes. For TEK simulations, we show two cases: with particle refill and without it. Here refill is a numerical scheme where the weights of those markers touching the radial computational boundary are set to zero and the markers are re-inserted at the mirror (to the midplane) poloidal location, while in the no-refill scheme particles touching the radial boundary are lost and not longer considered. The two schemes give similar TAE frequency but significantly different growth rate, with the refill scheme giving about 20% higher growth rate.

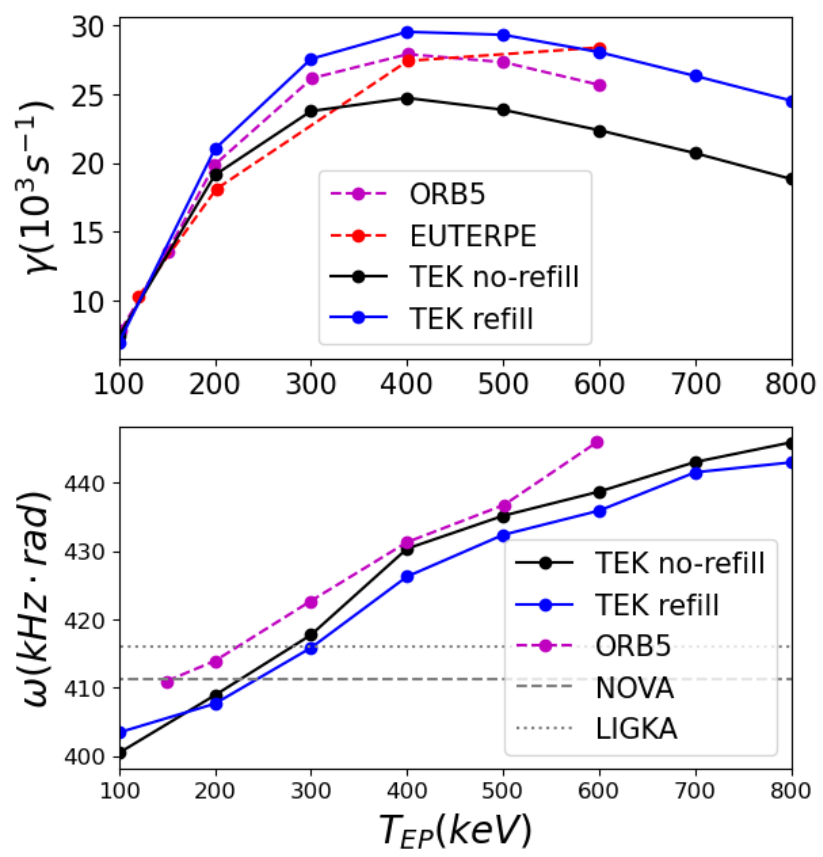

**Figure 8.** Dependence of the $n=6$ TAE growth rate $\gamma$ and angular frequency $\omega$ on the EP temperature. Two different results form `TEK` simulations are shown: one using the refill scheme and one without using it. As a reference, frequency calculated by two eigenvalues codes (`NOVA`[20] and `LIGKA`) is also shown, which is independent of the EP temperature. All data other than `TEK` results are from Ref. [3].

The different growth rates given by the refill and no-refill schemes affect the nonlinear saturation, which are discussed next. Unless stated otherwise, the refill scheme is used in getting the results presented in the paper.

## 4 Nonlinear simulation

Next, we consider nonlinear simulations. We first consider the single-$n$ ($n=6$) simulation. Then we consider the multiple-$n$ simulation that includes both $n=0$ and $n=6$ modes, and compare it with Liu Chen's analytical theory on zonal field beat-driven by Alfven eigenmodes[4].

### 4.1 Single $n=6$ simulation

In the single-$n$ simulation, we filter $\delta\Phi$ and $\delta A_{\|}$ to retain only the $n=6$ harmonic in each time step before the field is used to push markers. Figure 9 plots the time evolution of $\delta\Phi$ and $\delta A_{\|}$ on the low field side. Both the evolution in the full radial range and at a single radial location are plotted. The results show a typical pattern in which the mode first grows and then saturates and decays.

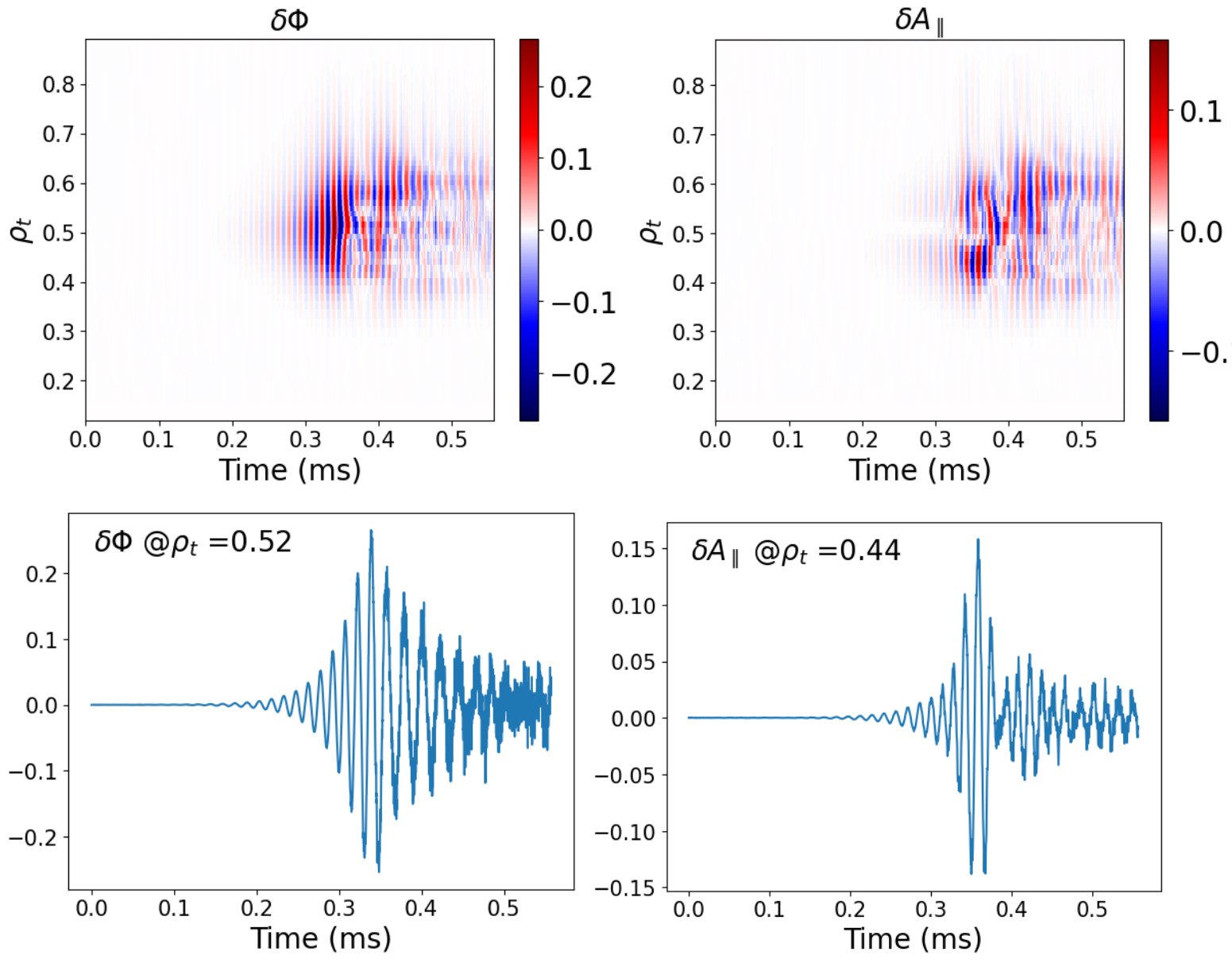

**Figure 9.** Time evolution of $\delta\Phi$ and $\delta A_{\parallel}$ on the low field side ($\theta = 0$, $\phi = 0$). Upper panel: in full radial range. Lower panel: at a single radial location (where the signal has the largest instantaneous value).

Figure 10 plots the continuous wavelet transform of $\delta\Phi(t)$ at $\rho_t = 0.52$. The results show that the frequency chirps down in the nonlinear stage.

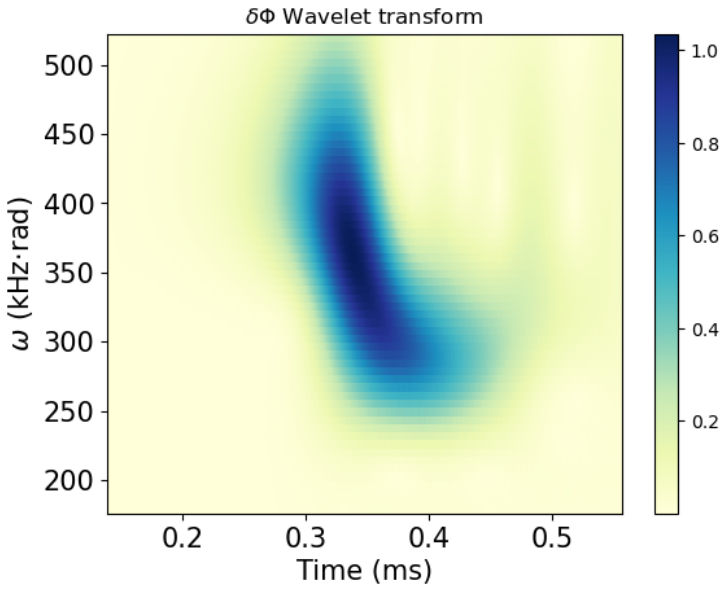

**Figure 10.** Continuous wavelet transform of $\delta\Phi(t)$ at $\rho_t = 0.52$. The transform was performed by using PyWavelets[21].

Figure 11 shows the time evolution of EP heat and particle flux, where the particle flux is defined by

$$Q_p = \left\langle \int \delta \mathbf{V}_D \cdot \frac{\nabla \psi}{|\nabla \psi|} \delta f d^3 \mathbf{v} \right\rangle_s , \tag{4}$$

and the heat flux is defined by

$$Q_h = \left\langle \int \delta \mathbf{V}_D \cdot \frac{\nabla \psi}{|\nabla \psi|} \left( \frac{1}{2} m_f v_\parallel^2 + m_f \mu B_0 \right) \delta f d^3 \mathbf{v} \right\rangle_s . \tag{5}$$

where $\langle \ldots \rangle_s$ is the magnetic surface averaging, $\delta f$ is the EP distribution function perturbation, and the other quantities are defined in Appendix A.2.

Figure 11 shows both radially resolved and the volume averaged values of the flux. The results indicate that the heat flux and particle flux show very similar time evolution, indicating EP kinetic energies do not change significantly. Radially, the peak flux appear near $\rho_t = 0.5$, where the TAE mode locates, indicating the flux is due to the TAE perturbation. The peak value of the volume-averaged heat flux is about $3\text{kW}/m^2$, which is negligibly small from reactor perspective. After the first major peak, the flux drops sharply to a very low value, and then resurges, forming a second minor peak, and then drops again to a very low value.

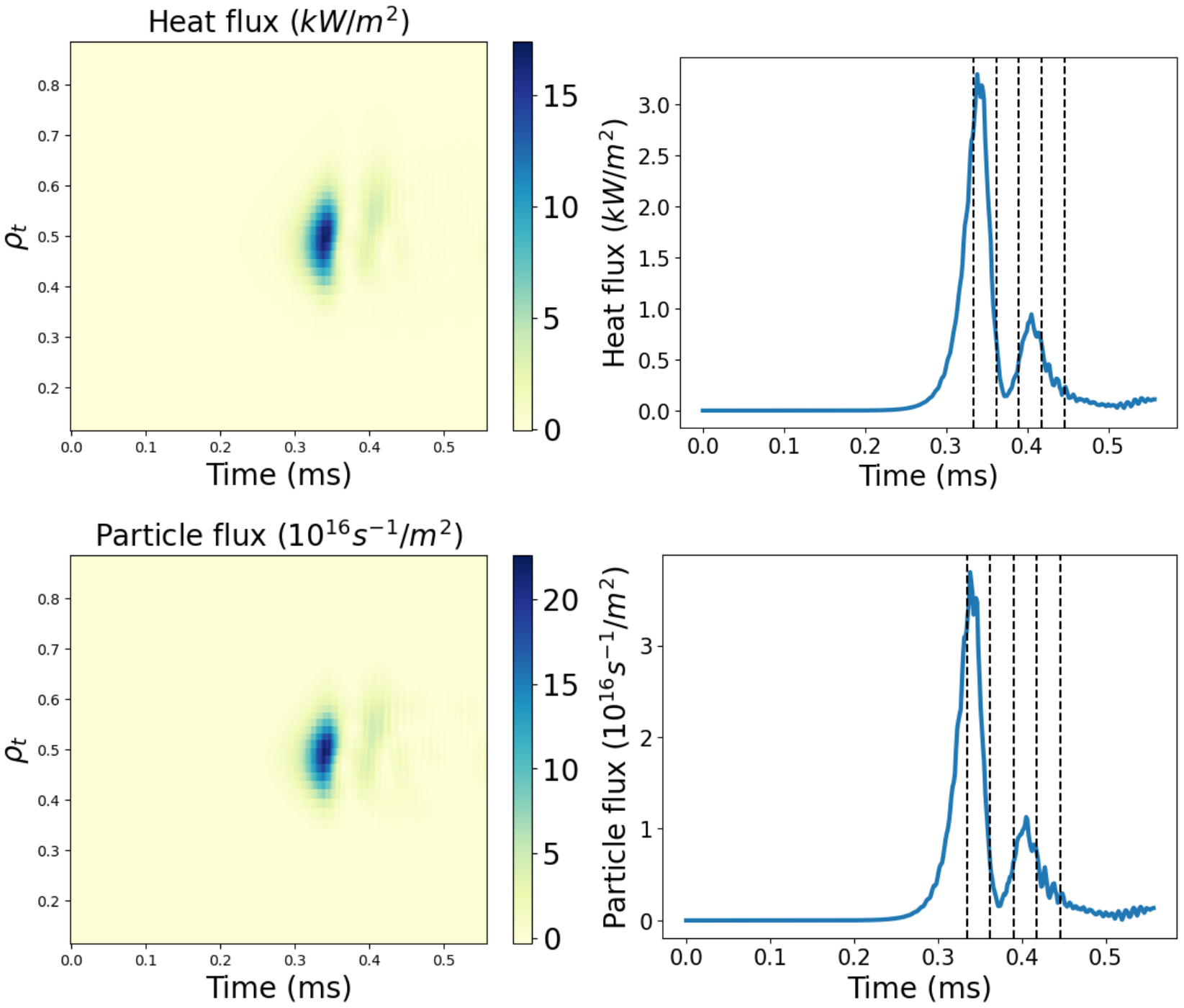

**Figure 11.** Time evolution of EP heat flux (upper panel) and particle flux (low panel) in single $n = 6$ nonlinear simulation. Left panel gives the radially resolved values. Right panel gives the volume averaged values. The vertical dashed lines mark the time slices at which the mode structures are plotted in Figs. 13 and 14.

Figure 12 compares the volume averaged EP heat flux obtained by using the refill and no-refill schemes. As is mentioned above, the higher linear growth rate given by the refill scheme has consequence on the nonlinear simulations. The general observation from Fig. 12 is that the nonlinear saturation is reached earlier and the saturation level is higher than the no-refill case. Except this difference, the time evolution is similar: after the first major peak, the flux drops sharply to a very low value, and then resurges, forming a second minor peak, and then drops again to a very low value.

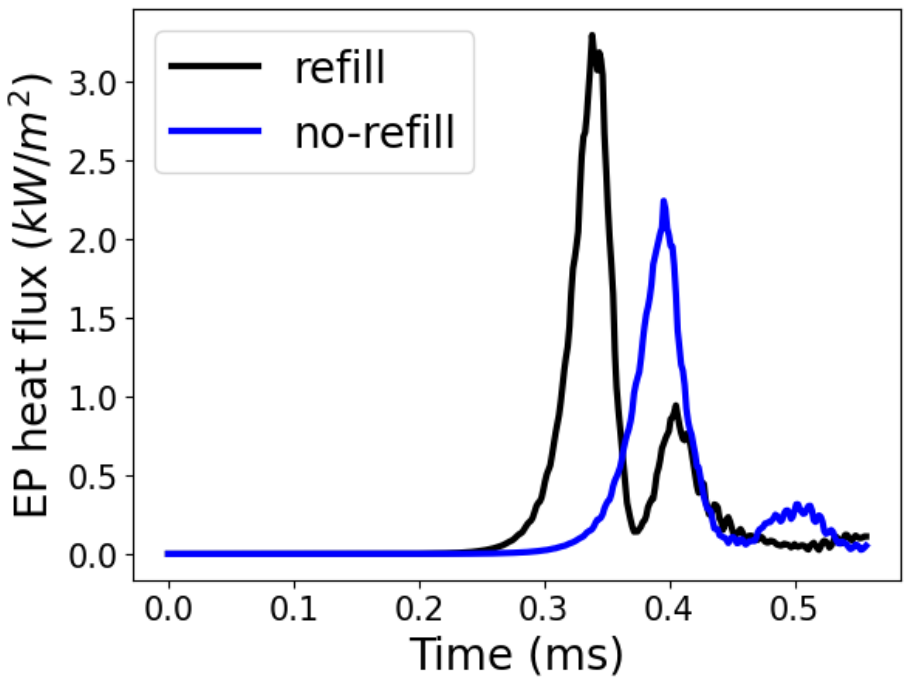

**Figure 12.** Comparison of volume averaged EP heat flux between the refill and no-refill for the single $n = 6$ simulation.

Next, let us examine snapshots of $\delta\Phi$ and $\delta A_{\|}$ in a poloidal plane, which are shown in Figure 13. The time slices selected, which are marked in Fig. 11, are roughly corresponding to the time when the heat flux reaches its peak ($t=0.3343$ms), drops sharply ($t=0.3621$ms), around the second minor peak ($t=0.3900$ms and $t=0.4178$ms), and drops again ($t=0.4457$ms), respectively.

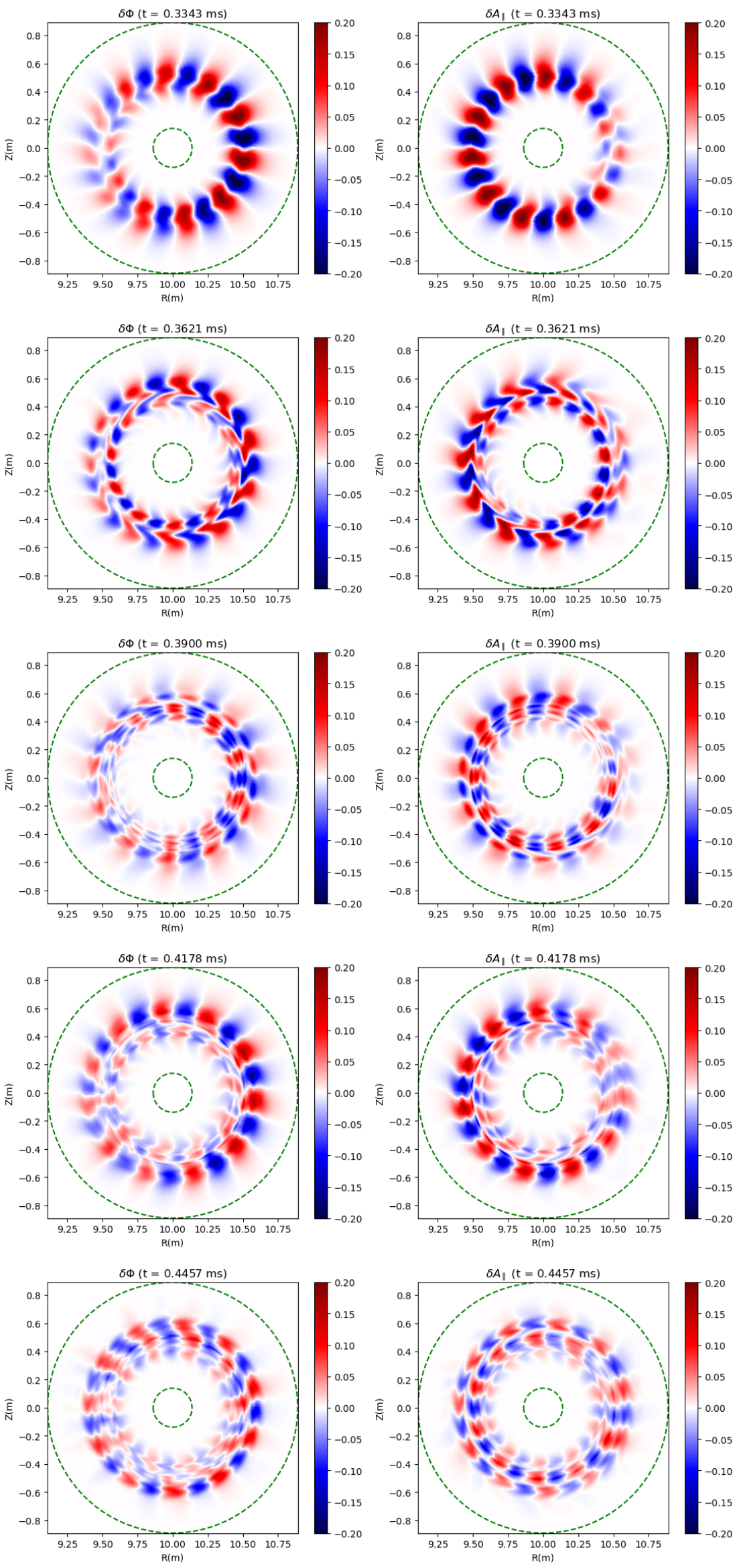

**Figure 13.** Snapshots of $\delta\Phi$ and $\delta A_{\|}$ in $\phi=0$ poloidal plane at different time slices.

Figure 14 plots radial profiles of amplitudes of $m=9,10,11,12$ poloidal harmonics. The time slices are corresponding to those in Fig. 13.

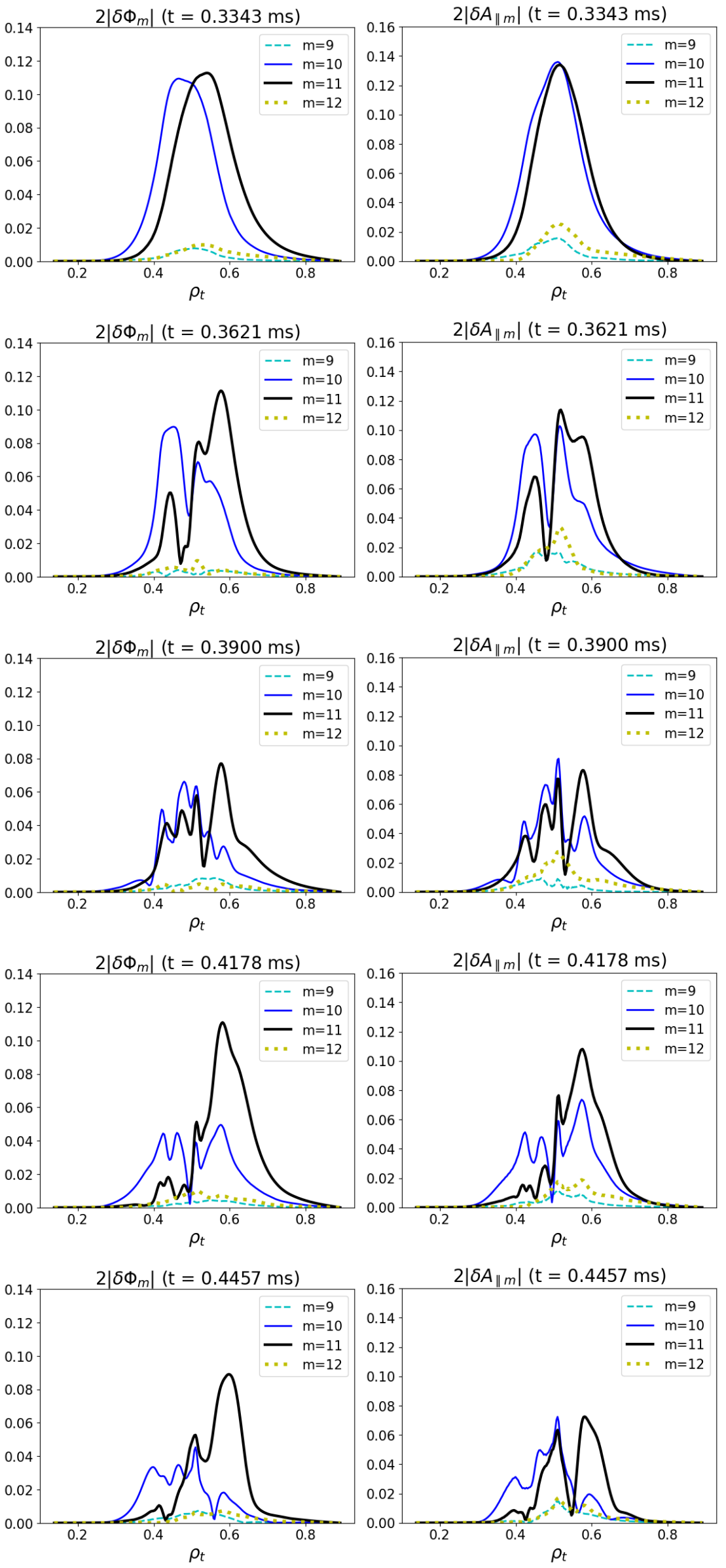

**Figure 14.** Radial profiles of amplitudes of $m = 9, 10, 11, 12$ poloidal harmonics.

A general observation from Figs. 13 and 14 is that small radial structures appear during the nonlinear stage. As is shown in Fig. 13, large structure of the 2D mode structure breaks down into small ones. The 1D amplitudes of poloidal harmonics in Fig. 14 show similar behavior: higher radial wavenumber structures appear. Another factor that has profound impact on the mode structure is that amplitude of the dominant poloidal harmonics ($m = 10, 11$) around $\rho_t = 0.5$ can decrease to nearly zero in the evolution, making the mode structure look as split in the radial direction. For the same amplitude, the finer radial structure here is less efficient in transporting EPs than the large one. This can be seen from the sharp drop in the EP heat flux when finer radial structure appears while the amplitude is not significantly reduced. The results in Figs. 13 and 14 also show that the transition between large and small radial structure is not one-way: finer structure can merge to form smooth one. The second minor peak in the EP flux, where mode structure is in a more smooth state, is related to this back transition, which results in flux increasing.

As to what mechanism is making the TAE saturate, decay, and sometime resurge, we do not have an answer. Exploring the physics behind this is beyond the scope of this paper.

Next, let us examine the EP density profile evolution, which is shown in Fig. 15. The change in the profile is so small that it is invisible in the profile plot. To show details, the right panel plots the difference between the equilibrium density profile and the instantaneous profiles.

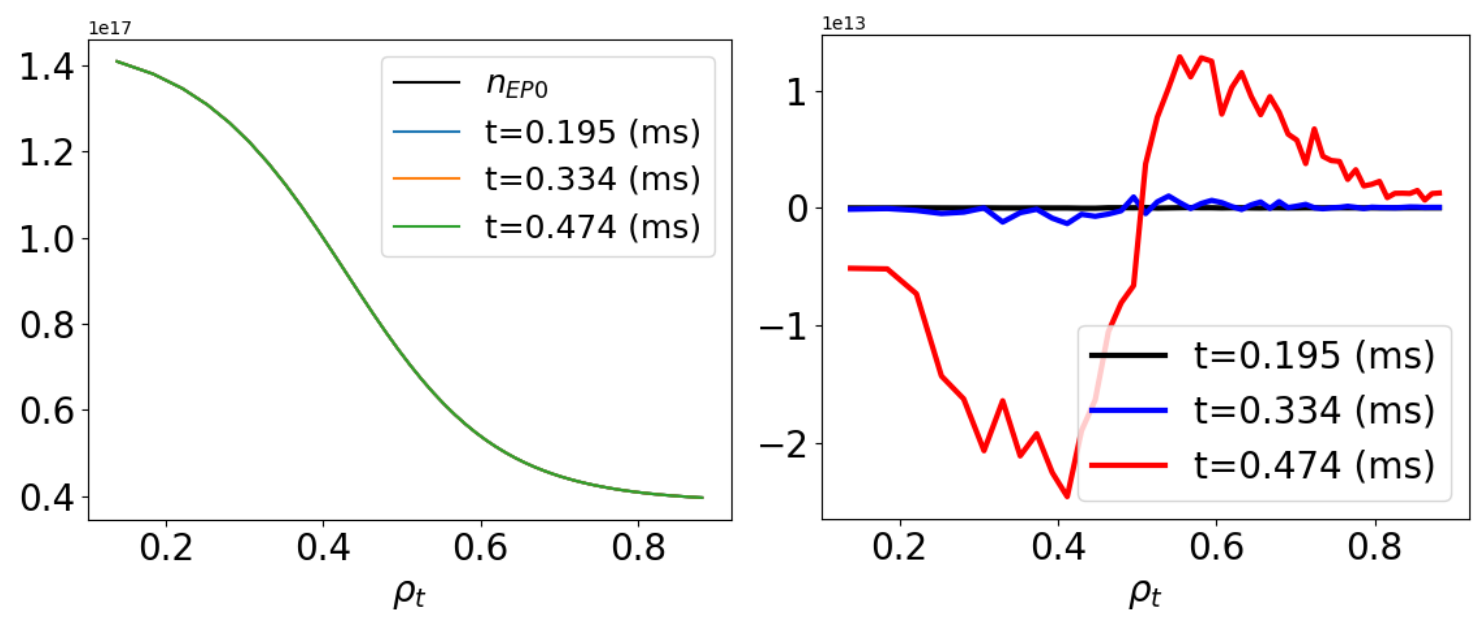

**Figure 15.** Left panel: Fast ion density profile at various time in the simulation. The change is so small that it is invisible. Right panel: the difference between the equilibrium density profile and the profile at $t$, $\Delta n = n(t) - n_{\mathrm{EP0}}$. The refill scheme is turned off.

The results show that the density in $\rho_t < 0.5$ region is reduced and the density in $\rho_t > 0.5$ is increased, i.e., the profile is being flatten around $\rho_t = 0.5$. We note that the TAE is centered around $\rho_t = 0.5$. So the flatten can be due to TAE transporting EPs radially outward. The flattening effect is very small, with the density variation being 4 orders smaller than the equilibrium density, as is shown in Fig. 15. This is consistent with the small EP flux observed in the simulation (Fig. 11).

### 4.2 $n = 0, 6$ simulation

In the simulations that include both $n = 0$ and $n = 6$ harmonics, we filter the $n = 0$ harmonic by retaining only its zonal component (i.e., its magnetic surface average). Without this filtering, simulations are numerically unstable. To verify that the zonal component are correctly simulated, we compare simulations with those predicted by the theory (see Fig. 20 )).

Figure 16 plots the time evolution of $\delta\Phi$ and $\delta A_{\|}$ on the low field side. Both the evolution in the full radial range and at a single radial location are plotted.

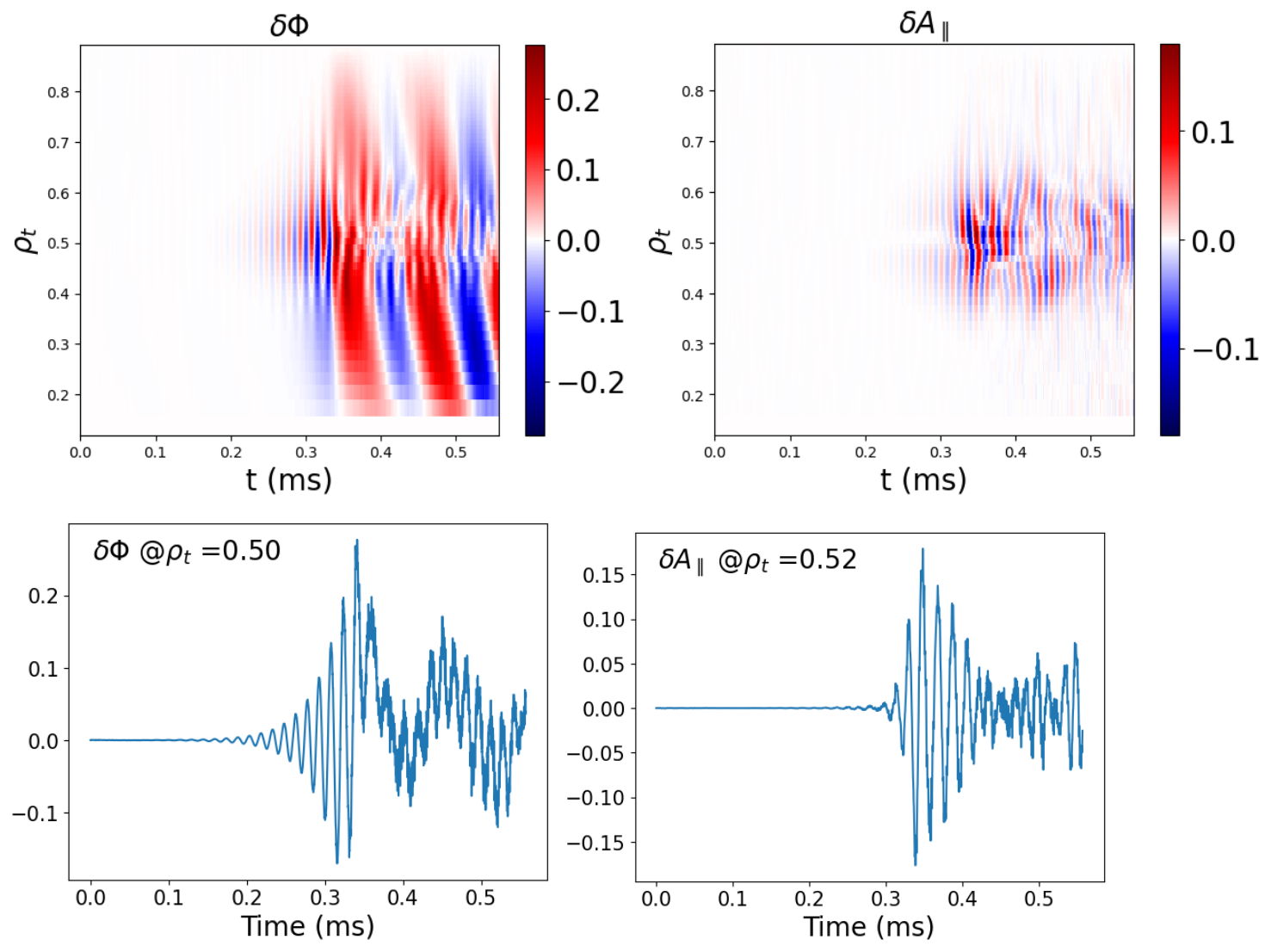

**Figure 16.** Time evolution of $\delta\Phi$ and $\delta A_{\|}$ in the low field side midplane ($\theta = 0, \phi = 0$). Upper panel: in full radial range, Lower panel: at a single radial location (the location selected is where the maximal instantaneous value is reached).

To see the frequency variation during the nonlinear evolution, Figure 17 plots the continuous wavelet transforms of $\delta\Phi(t)$ and $\delta A_{\|}(t)$ at chosen radial locations. For $\delta\Phi$, we see the frequency chirps down and there are low-frequency modes accompanying the TAE mode, but for $\delta A_{\|}$ there is no low-frequency mode appearing.

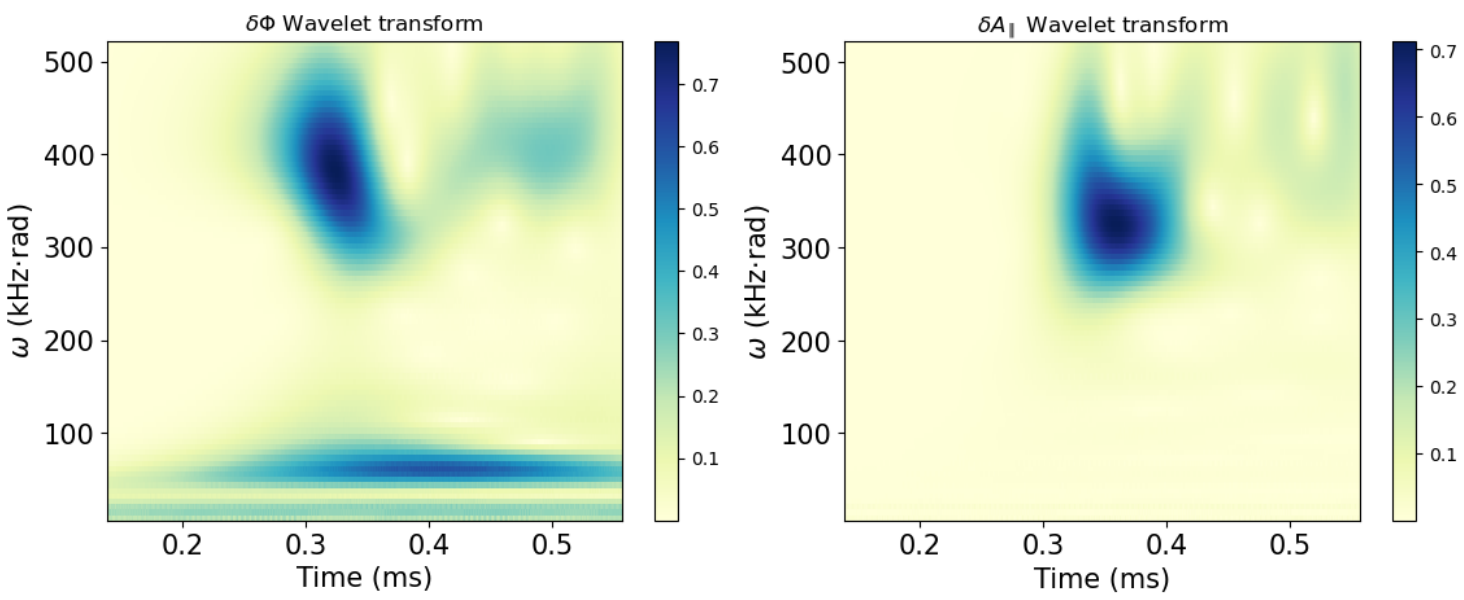

**Figure 17.** Continuous wavelet transforms of $\delta\Phi(t)$ at $\rho_t = 0.50$ (left), and $\delta A_\parallel(t)$ at $\rho_t = 0.52$ (right).

Figures 18 and 19 decompose the perturbations into $n=0$ and $n=6$ harmonics, in order to compare their amplitudes. Both the evolution in the full radial range and at a single radial location are plotted. For $\delta\Phi$, the amplitude of $n=0$ harmonic is larger than that of the $n=6$ harmonic, but for $\delta A_\parallel$, the former is two orders smaller than the latter.

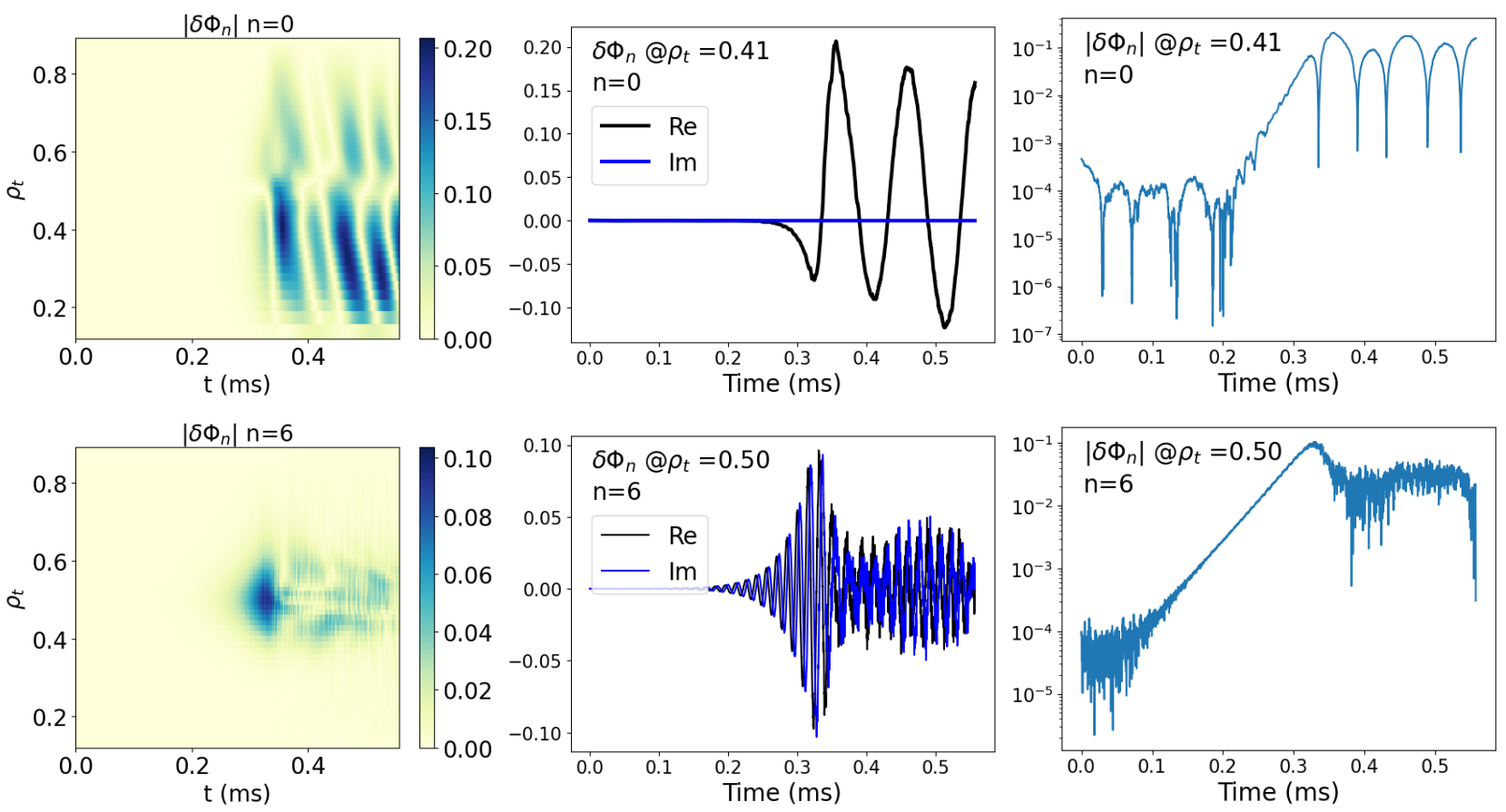

**Figure 18.** Time evolution of the $n=0$ (upper panel) and $n=6$ (lower panel) toroidal harmonics of $\delta\Phi$ on the low field side ($\theta = 0$).

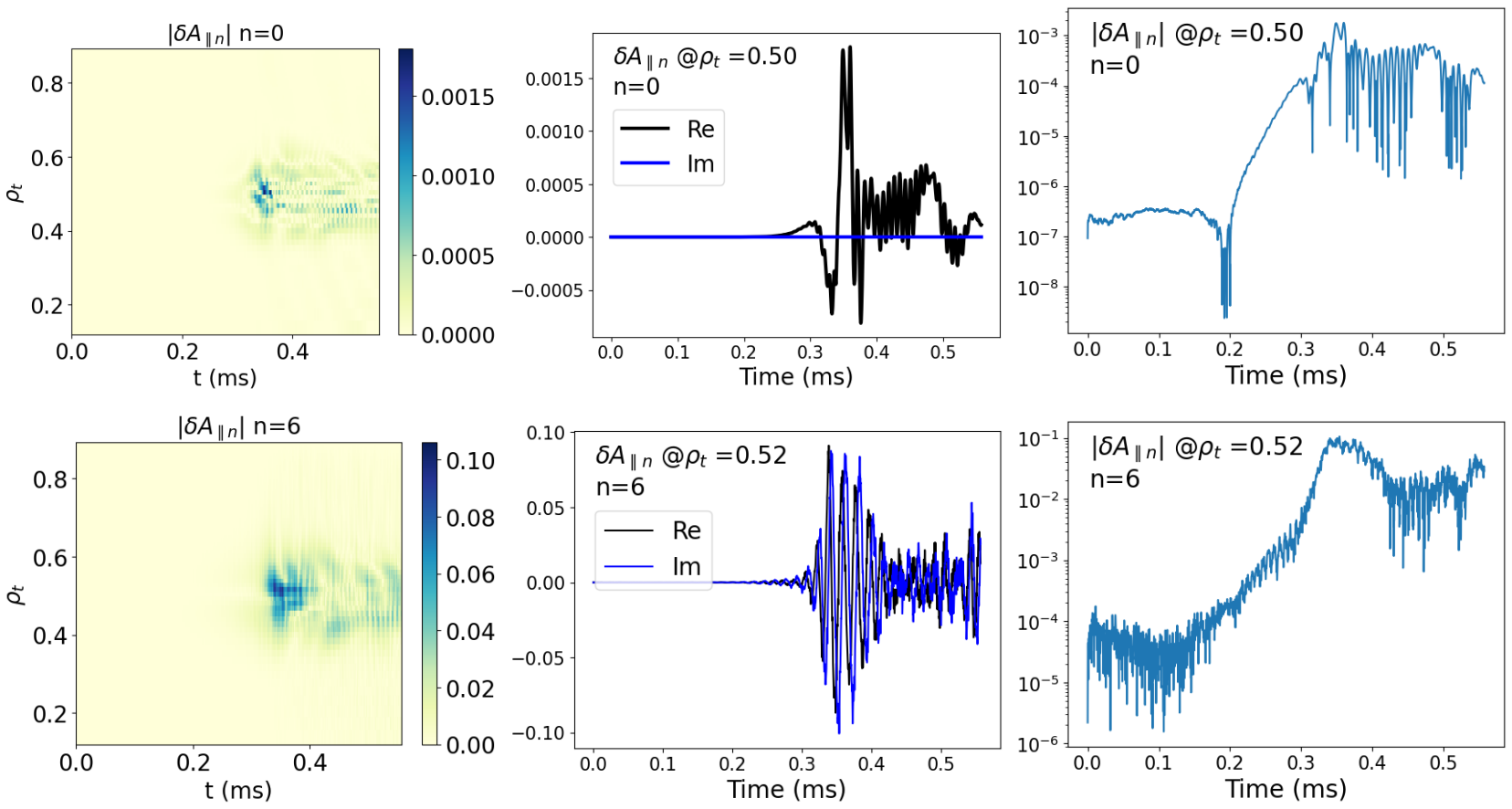

**Figure 19.** Time evolution of the $n=0$ (upper panel) and $n=6$ (lower panel) toroidal harmonics of $\delta A_\parallel$ on the low field side ($\theta=0$).

Next, we compare the zonal current $\delta A_{\|z}$ obtained in simulations with those predicted by the theory developed in Ref. [4]. The zonal current is defined by $\delta A_{\|z}=\langle\delta A_\|\rangle_s$ where $\langle\ldots\rangle_s$ is the magnetic surface averaging. The theory provides a formula for calculating $\delta A_{\|z}$ from the main Alfven eigenmodes:

$$\delta A_{\|z}=\frac{1}{B_{0a}}\frac{\partial}{\partial r}\sum_m\left(\frac{nq}{r}\right)\left(\frac{m-nq}{qR_0}\right)^{-1}|\delta A_{\|m}|^2, \tag{6}$$

where $B_{0a}$ is the magnetic field strength at the magnetic axis, $\delta A_{\|m}$ is the poloidal Fourier expansion coefficient of the TAE. This theory is semi-analytic: it relies on simulations to provide the nonzero-$n$ ($n=6$ in this case) TAE mode structure and amplitude. The nonzero-$n$ mode is the main Alfven eigenmode that is expected to drive the zonal current $\delta A_{\|z}$.

Figure 20 compares the zonal current $\delta A_{\|z}$ at various time in the simulation with those predicted by the theory (obtained by using Eq. (6) with $n=6$). The results show that the simulation and theory are in agreement in both the radial shape and amplitude. The agreement is excellent considering that the agreement is achieved during the entire duration, where the amplitude has grew by two orders, and the process goes through the linear growth, nonlinear saturation, and decay phase. Even at the end of the simulation ($t = 0.56$ms), where the TAE amplitude has decayed to a very low value with finer radial structure, the agreement is still decent.

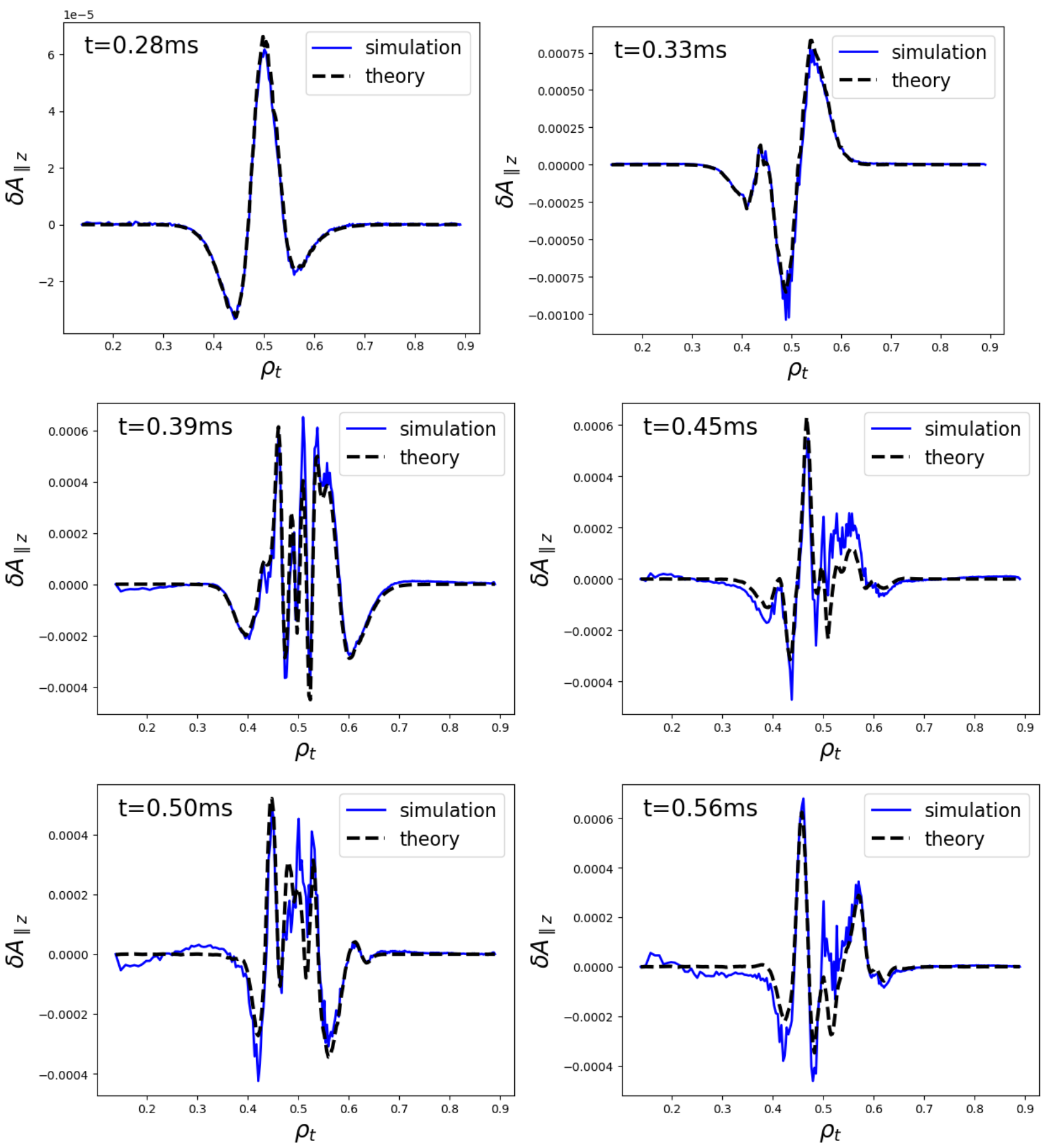

**Figure 20.** Zonal field $\delta A_{\|z}$ at various time observed in the simulation and those predicted by the theory, Eq. (6).

In order to show that the refill scheme does not affect the zonal field generation process, Fig. 21 plots the comparison between the simulation and theory in the no-refill scenario, which show similar good agreement as in the refill case.

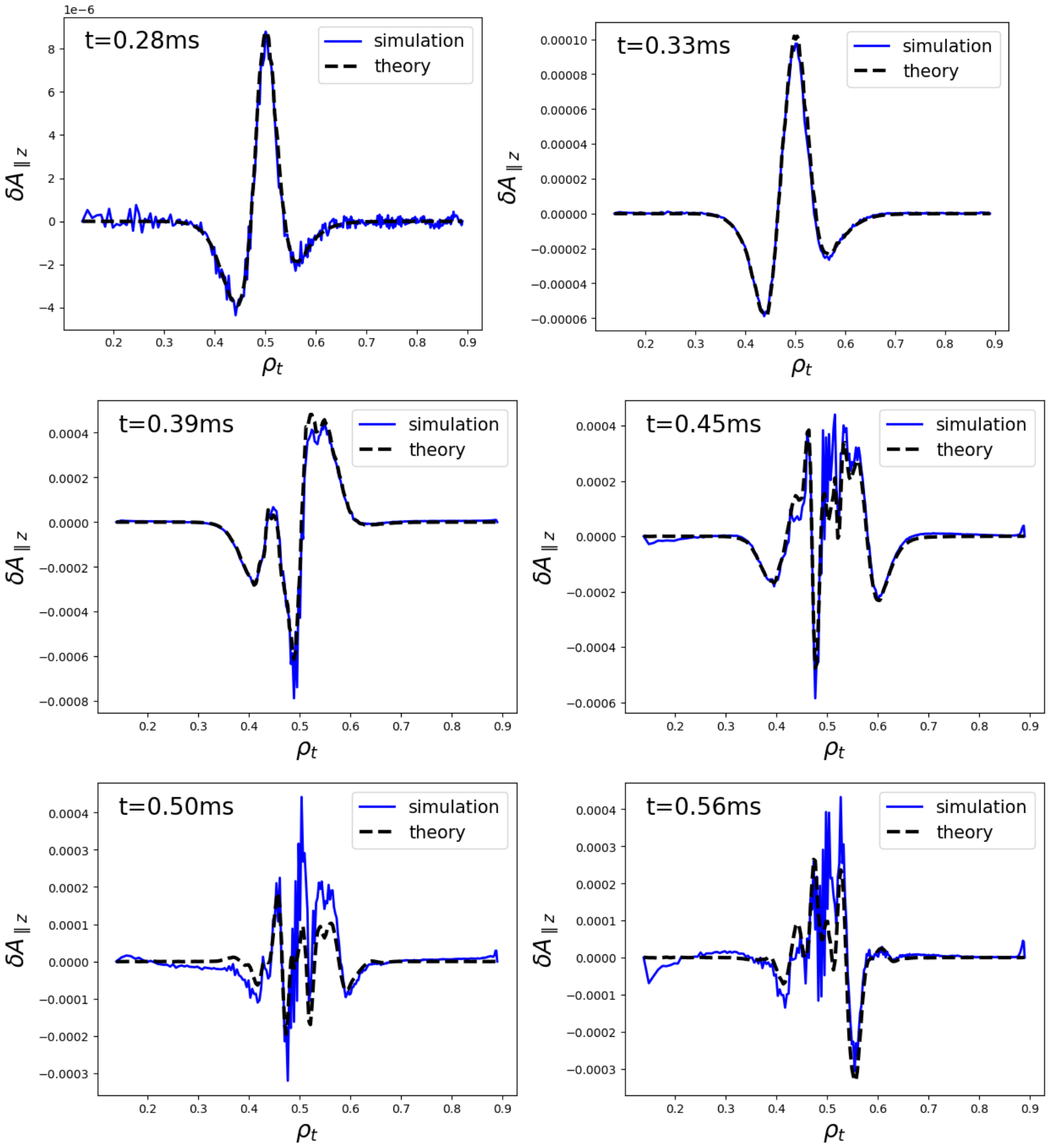

**Figure 21.** The same as Fig. 20 except that this is in the no-refill case.

Figure 22 plots snapshots of the $n = 6$ harmonic (i.e., excluding the $n = 0$ part from the signal) of $\delta\Phi$ and $\delta A_{\|}$ in the poloidal plane and the corresponding amplitude of the dominant poloidal harmonics.

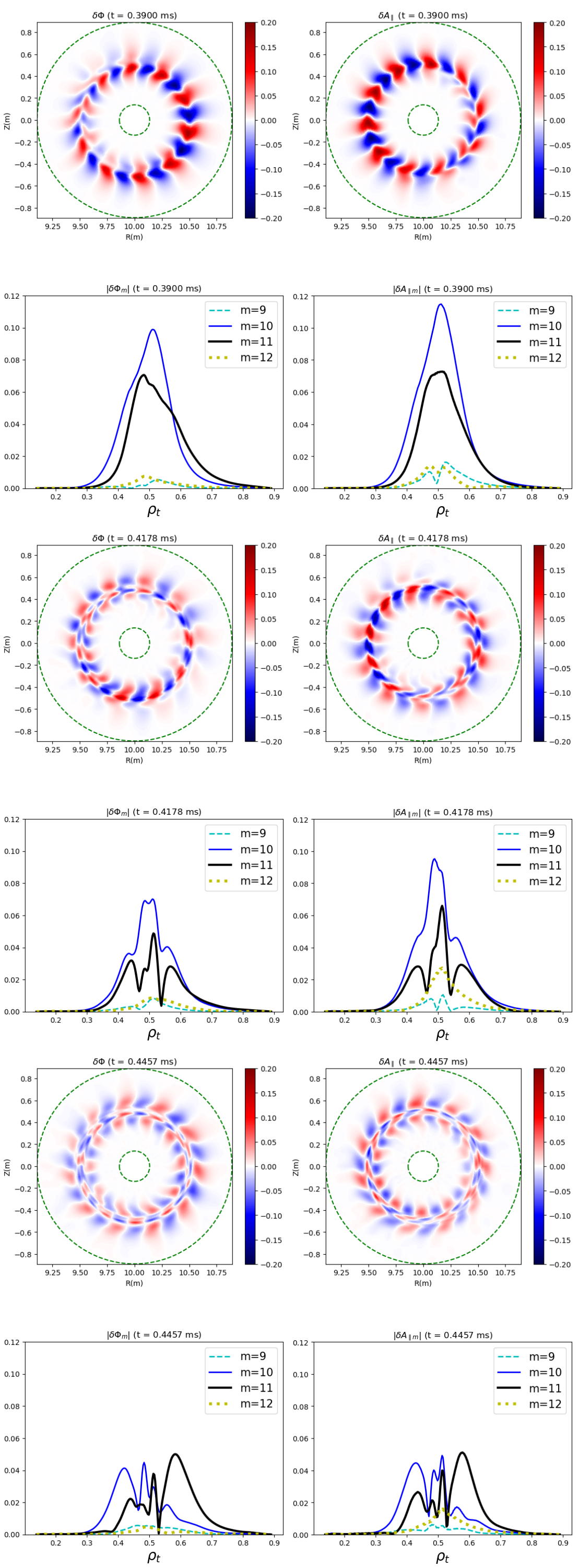

**Figure 22.** Snapshots of $n \neq 0$ part of $\delta\Phi$ and $\delta A_{\|}$ (namely the $n=6$ harmonic) in the poloidal plane and the corresponding amplitude of the dominant poloidal harmonics.

Figure 22 shows a nonlinear saturation pattern similar to that of the single $n=6$ simulation shown in Figs. 13 and 14. But the amplitude is smaller than that in Figs. 13 and 14. In this sense, the $n=0$ mode suppresses the $n=6$ TAE. The reduction in the TAE amplitude also manifests itself in the reduction in the EP heat flux, which is plotted in Fig. 23. The peak heat flux is reduced by 47% from that in the single $n=6$ simulation (Fig. 11).

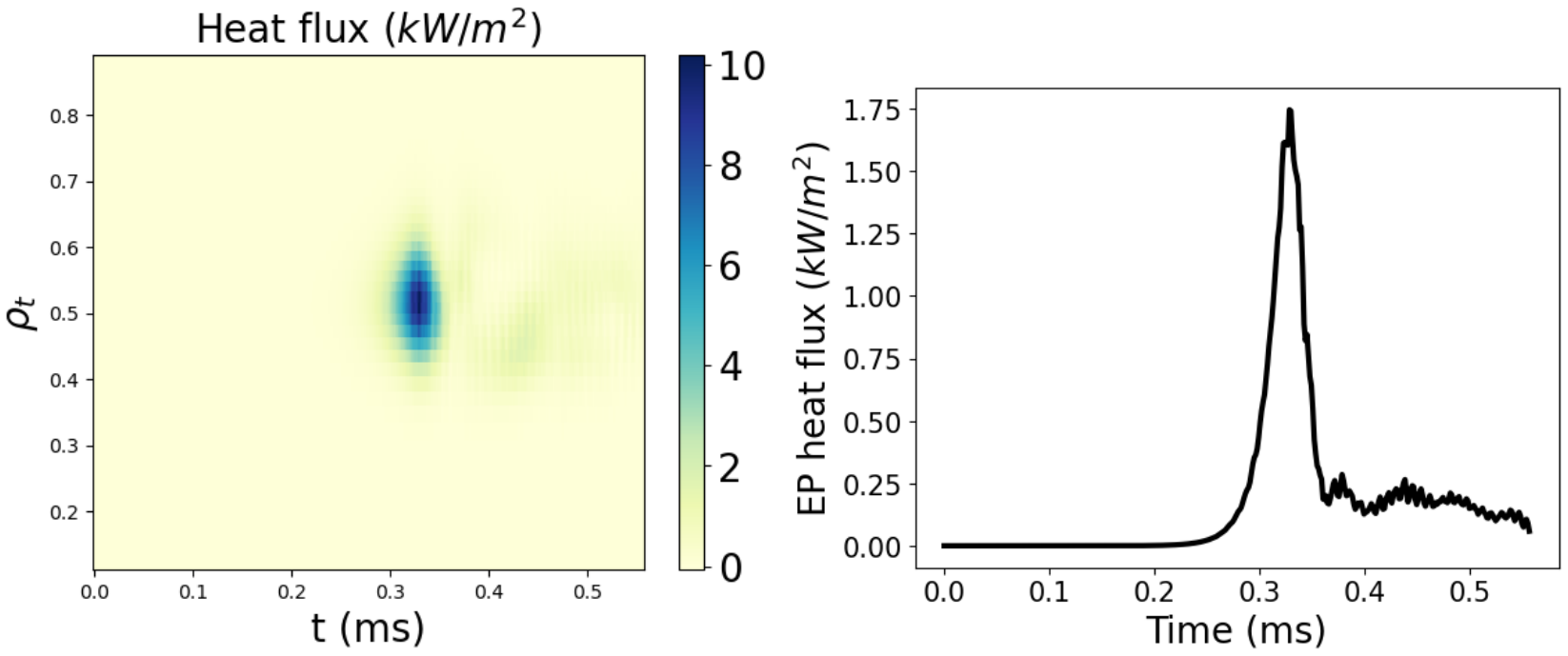

**Figure 23.** Time evolution of EP heat flux in multiple-$n$ ($n=0,6$) simulation. Left panel gives the radially resolved value. Right panel gives the volume averaged value.

Figure 24 compares the volume averaged EP heat flux obtained by using the refill and no-refill schemes. The general observation from Fig. 24 is similar to that from Fig. 12: in the refill case, the nonlinear saturation is reached earlier and the saturation level is higher than the no-refill case. Except this difference, the time evolution is similar: after the first major peak, the flux drops to a very low value.

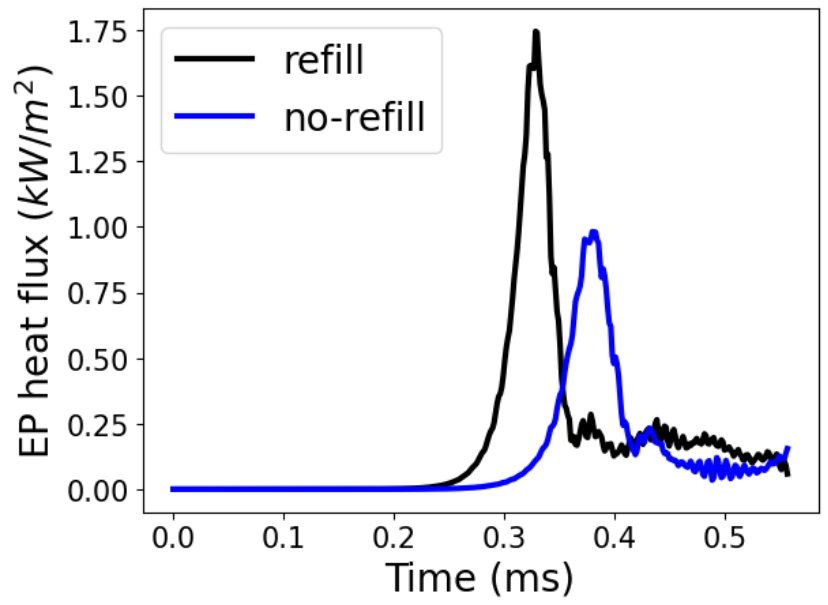

**Figure 24.** Comparison of volume average EP heat flux between the refill and no-refill.

Figure 25 plots the EP density profiles and density perturbations at various time slices. The results are similar to those in the single $n=6$ simulation shown in Fig. 15 (except that the density perturbation is only half of that in Fig. 15).

The conclusions got there are also valid here.

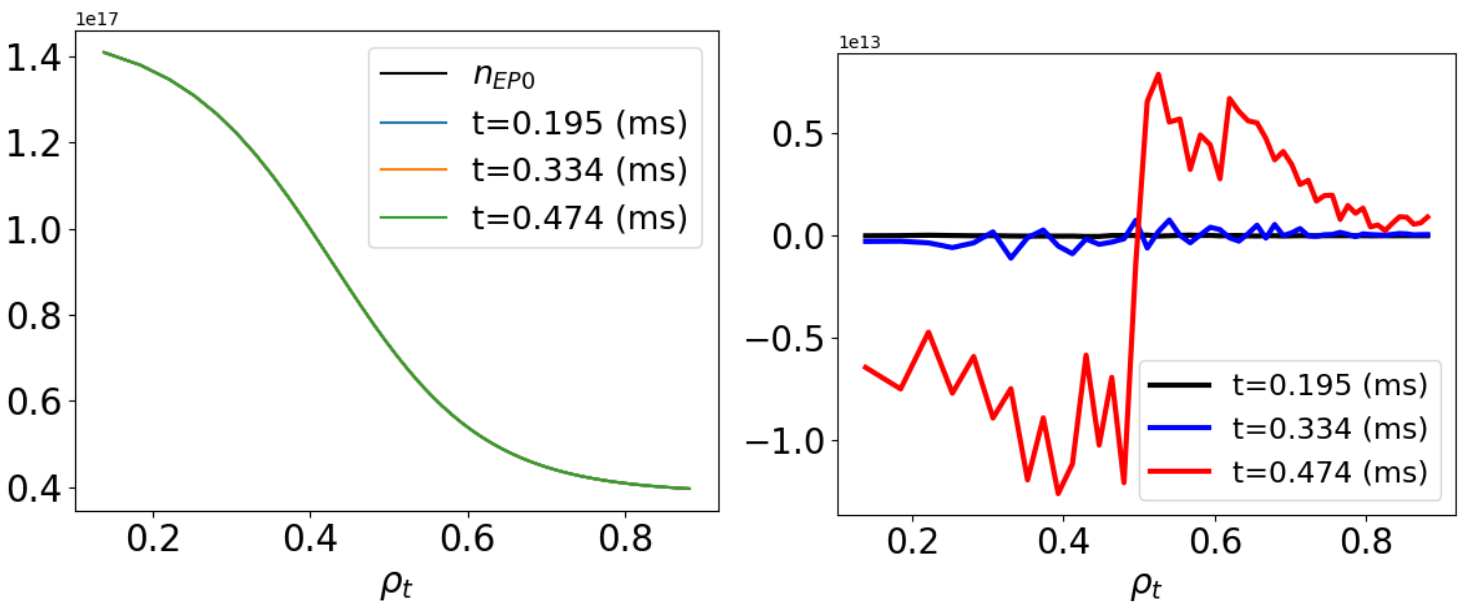

**Figure 25.** Fast ion density profile (left) and perturbation (right) at various time slices in the simulation. The refill scheme is turned off.

# 5 Summary and discussion

Linear and nonlinear gyrokinetic simulations in the ITPA-EP TAE benchmark case were performed using a new $\delta f$ PIC code named `TEK`. In the linear simulations, the $n=6$ TAE mode structure are examined in details, clarifying the reason why different codes have discrepancy in the radial structure of the poloidal harmonics: they are using different definitions of the poloidal angle.

In the nonlinear simulations, both single-$n$ ($n=6$) and multiple-$n$ ($n=0$, 6) simulations are presented. A comparison between simulations and theory indicates the zonal component of $\delta A_\parallel$ (often called zonal current) is correctly simulated. A conclusion from the nonlinear simulations is that the TAE induces very small transport of EPs, with the volume-averaged peak value of the EP heat flux being about 3kW$/m^2$ in the single-$n$ case and 2kW$/m^2$ in the multiple-$n$ case. Furthermore, this flux quickly drops to nearly zero level after the peak. As a result the EP profiles remain almost unchanged. The peak value of volume-averaged EP heat flux is an important quantity that should be compared in future inter-code benchmarking.

Besides the $n=6$ TAE, we found $n=3,4$ modes are also driven unstable by the EPs in this configuration (not shown in this paper), but have not been compared between codes. So EP physics in this simple configuration is still open to further investigation. We note that the plasma equilibrium used in the TAE benchmarking is not a configuration based on actually experiments (unlike the DIII-D cyclone base case that is based on experiments). However many researcher have invested time in doing simulations in this configuration and accumulated results that might be used as benchmarks. Considering this, it is still interesting to do detailed simulations in this configuration and compare results between codes.

# 6 Acknowledgments

Y. Hu would like to thank Dr. Baobao Jia for useful discussion on GTC simulations. This material is based on work supported by National Key R&D Program of China under Grant No. 2024YFE03170003, Strategic Priority Research Program of Chinese Academy of Sciences under Grant No. XDB0790203, and the National Natural Science Foundation of China under Grant No. 12475234. Numerical computations of this work were performed on machines in Tianjin Tianhe computing center and Sugon Hefei ChaohuMingYue.

# A Maxwell's equations, gyrokinetic equation, coordinates, and numerical algorithm used in TEK

## A.1 Maxwell's equations

Using the scalar potential $\delta\Phi$ and vector potential $\delta\mathbf{A}$, Maxwell's equations in the Coulomb gauge (i.e., $\nabla\cdot\delta\mathbf{A}=0$) with the displacement current neglected (because we are considering low-frequency modes) are written as

$$-\nabla^2\delta\mathbf{A}=\mu_0\delta\mathbf{J}, \tag{7}$$

$$-\nabla^2\delta\Phi=\frac{\delta\rho_c}{\varepsilon_0}, \tag{8}$$

where $\mu_0$ and $\varepsilon_0$ are the vacuum magnetic permeability and vacuum permittivity, respectively, $\delta\mathbf{J}$ and $\delta\rho_c$ are the current density and charge density, respectively. The electric field $\delta\mathbf{E}$ and magnetic field $\delta\mathbf{B}$ are given by

$$\delta\mathbf{B}=\nabla\times\delta\mathbf{A}, \tag{9}$$

$$\delta\mathbf{E}=-\nabla\delta\Phi-\frac{\partial\delta\mathbf{A}}{\partial t}. \tag{10}$$

A further approximation adopted in this work and most gyrokinetic codes (e.g. GEM[22], NLT[23, 24], GKNET[25]) is to keep only the parallel component of $\delta\mathbf{A}$, i.e., $\delta\mathbf{A}\approx\delta A_{\|}\mathbf{e}_{\|}$, where $\mathbf{e}_{\|}=\mathbf{B}_0/B_0$. Then Ampere's law (7) is approximated as

$$-\nabla^2\delta A_{\|}=\mu_0\delta J_{\|}, \tag{11}$$

where$\delta J_{\|}=\delta\mathbf{J}\cdot\mathbf{e}_{\|}$ is the parallel current density. In this case, only one of the three components of $\delta\mathbf{J}$, namely $\delta J_{\|}$, needs to be computed.

## A.2 Gyrokinetic equation

The gyrokinetic equation used in TEK is a reformulation of the Frieman-Chen form [26, 27], which, for an isotropic equilibrium distribution $F_0$, is given by

$$\begin{aligned}&\frac{\partial\delta G_0}{\partial t}+(v_{\|}\mathbf{e}_{\|}+\mathbf{V}_{D0}+\delta\mathbf{V}_D)\cdot\nabla\delta G_0\\&=-\delta\mathbf{V}_D\cdot\nabla F_0-\frac{q}{m}\frac{\partial\langle\delta L\rangle}{\partial t}\frac{\partial F_0}{\partial\varepsilon},\end{aligned} \tag{12}$$

where $\delta G_0 = \delta G_0(\mathbf{X}, \varepsilon, \mu, t)$, $\mathbf{X}$ is the guiding-center position, $\varepsilon = v^2/2$, $\mu = v_\perp^2/2B_0$; $\delta G_0$ is related to the distribution function perturbation $\delta f$ by

$$\delta f = \frac{q}{m}\delta\Phi\frac{\partial F_0}{\partial\varepsilon} + \delta G_0, \tag{13}$$

where the first term is called adiabatic term, $q$ is the species charge (rather the safety factor used in the main text), $m$ is the species mass (rather than the poloidal mode number), $\langle\ldots\rangle$ is the gyro-phase averaging operator; $\delta L = \delta\Phi - \mathbf{v}\cdot\delta\mathbf{A} \approx \delta\Phi - v_\parallel\delta A_\parallel$; $\mathbf{V}_{D0}$ is the guiding-center drift velocity in the equilibrium field, and $\delta\mathbf{V}_D \equiv -\frac{q}{m}\nabla\langle\delta L\rangle\times\frac{\mathbf{e}_\parallel}{\Omega}$ is the drift velocity perturbation, which consists of the $\delta\mathbf{E}\times\mathbf{B}_0$ drift and magnetic fluttering, where $\Omega = B_0 q/m$.

The Frieman-Chen equation (12) contains time derivatives of unknown $\delta\Phi$ and $\delta\mathbf{A}$ on the right-hand side, which is problematic from numerical perspective (e.g., numerical instabilities can arise if discretized by explicit finite difference schemes). So many authors prefer to reformulate it by eliminating these time derivatives. One popular way to achieve this is to split $\delta f$ as[28]

$$\delta f = \delta f^{(p_\parallel)} + \underbrace{\frac{q}{m}(\delta\Phi - \langle\delta\Phi\rangle)\frac{\partial F_0}{\partial\varepsilon}}_{\text{term 1}} + \underbrace{\frac{q}{m}\langle v_\parallel\delta A_\parallel\rangle\frac{\partial F_0}{\partial\varepsilon}}_{\text{term 2}}, \tag{14}$$

then $\delta f^{(p_\parallel)}$ satisfies the following gyrokinetic equation:

$$\begin{aligned} & \left[\frac{\partial}{\partial t} + (v_\parallel\mathbf{e}_\parallel + \mathbf{V}_{D0} + \delta\mathbf{V}_D)\cdot\nabla\right]\delta f^{(p_\parallel)} \\ = \; & -\delta\mathbf{V}_D\cdot\nabla F_0 + \frac{q}{m}\frac{\partial F_0}{\partial\varepsilon}(v_\parallel\mathbf{e}_\parallel + \mathbf{V}_{D0})\cdot\nabla\langle\delta\Phi\rangle \\ - \; & \frac{q}{m}\frac{\partial F_0}{\partial\varepsilon}[v_\parallel(v_\parallel\mathbf{e}_\parallel + \mathbf{V}_{D0})\cdot\nabla\langle\delta A_\parallel\rangle - \langle\delta A_\parallel\rangle\mu\mathbf{e}_\parallel\cdot\nabla B_0], \end{aligned} \tag{15}$$

which is the so-called $p_\parallel$ formulism, in which no time derivatives of $\delta\Phi$ or $\delta A_\parallel$ appear on the right-hand side.

The zeroth order moment of term 1 and the $v_\parallel$ moment of term 2 in Eq. (14) give rise to the polarization density and the skin current, respectively. The polarization density turns out helpful when we solve the Poisson equation, but the skin current turns out to be problematic when we solve the parallel Ampere's law. The skin current $\delta j_\parallel^{(\text{skin})}$, which is equal to $-(n_0 q^2/m)\delta A_\parallel$ for Maxwellian $F_0$ in the zero Larmor radius limit, is proportional to plasma density $n_0$. For high density plasma, gyrokinetic simulations indicate the skin current can be several order larger than the physical current carried by $\delta f$. This means that $\delta j_\parallel^{(\text{skin})}$ nearly cancels the current carried by $\delta f^{(p_\parallel)}$, giving a small net current. This raises the question: is numerical cancellation error significant? It turns out that this error is indeed significant, which gives rise to numerical instabilities if no special treatment is used.

There are several methods to mitigate the cancellation problem in the $p_\parallel$ formulism[27, 6]. One of the methods is the "mixed-variable pullback" method developed by Mishchenko et al [6]. This method is straightforward and easy to understand. In this method, we predict a guess value of $\delta A_\parallel$ using a time evolution equation and solve only the remainder from the Ampere equation. If our guess is good, then the remainder will be small so that the cancellation error will be small. How good the guess is of course depends on how good the evolution equation is (and how large the time step size is). At the end of each time step, we have a chance to calibrate our guess by using the correct value of $\delta A_\parallel$ as the initial value of the evolution equation. This avoid the error accumulation and confine the deviation from the guess within one time step. Specifically, we define $\delta A_\parallel^{(h)}$ by

$$\delta A_\parallel^{(h)} = \delta A_\parallel - \delta A_\parallel^{(s)}, \tag{16}$$

with $\delta A_\parallel^{(s)}$ determined by an evolution equation (inspired by the ideal Ohm's law, i.e. $\delta E_\parallel = 0$):

$$\frac{\partial \delta A_\parallel^{(s)}}{\partial t} = -\mathbf{e}_\parallel \cdot \nabla \delta\Phi. \tag{17}$$

Using $\delta A_\parallel^{(h)}$, we define a new distribution function $\delta f^{(h)}$ by

$$\delta f = \delta f^{(h)} + \frac{q}{m}(\delta\Phi - \langle\delta\Phi\rangle)\frac{\partial F_0}{\partial \varepsilon} + \frac{q}{m}\left\langle v_\parallel \delta A_\parallel^{(h)}\right\rangle \frac{\partial F_0}{\partial \varepsilon}. \tag{18}$$

(Comparing this with Eq. (14), we know the new skin current is proportional to $\delta A_\parallel^{(h)}$ rather than $\delta A_\parallel$. So if we manage to keep $\delta A_\parallel^{(h)}$ small, then we reduce the cancellation error.) Using this split of $\delta f$ into the Frieman-Chen equation, we obtain an equation for $\delta f^{(h)}$:

$$\begin{aligned}
&\left[\frac{\partial}{\partial t} + (v_\parallel \mathbf{e}_\parallel + \mathbf{V}_{D0} + \delta\mathbf{V}_D)\cdot\nabla\right]\delta f^{(\mathrm{mv})} \\
&= -\delta\mathbf{V}_D \cdot \nabla F_0 + \frac{q}{m}\frac{\partial F_0}{\partial\varepsilon}(\mathbf{V}_{D0} + \delta\mathbf{V}_D)\cdot\nabla\langle\delta\Phi\rangle \\
&-\frac{q}{m}\frac{\partial F_0}{\partial\varepsilon}\left[v_\parallel(v_\parallel \mathbf{e}_\parallel + \mathbf{V}_{D0} + \delta\mathbf{V}_D)\cdot\nabla\left\langle\delta A_\parallel^{(h)}\right\rangle - \left\langle\delta A_\parallel^{(h)}\right\rangle \mu \mathbf{e}_\parallel \cdot \nabla B_0\right].
\end{aligned} \tag{19}$$

Meanwhile, the parallel Ampere equation

$$-\nabla_\perp^2 \delta A_\parallel = \mu_0 \sum_j \delta J_{||j}, \tag{20}$$

is written as

$$\left(\left(\mu_0 \sum_j \frac{n_{0j} q_j^2}{m_j}\right) - \nabla_\perp^2\right)\delta A_\parallel^{(h)} = \nabla_\perp^2 \delta A_\parallel^{(s)} + \mu_0 \sum_j \delta J_{||j}^{(h)}, \tag{21}$$

where $\delta J_{\|j}^{(h)}$ is the parallel current carried by $\delta f_j^{(h)}$, the subscript $j$ is species index ($j = i, e, f$ for thermal ions, electrons, and EPs). Note that $\delta A_{\|}^{(s)}$ has been moved to the right-hand side as a source term because its value is already known (by integrating Eq. (17)) before we solve the Ampere equation for $\delta A_{\|}^{(h)}$. Here $F_0$ is taken to be Maxwellian and zero Larmor radius is assumed for both ions and electrons in getting the first term: $\mu_0 \delta A_{\|}^{(h)} n_{0j} q_j^2 / m_j$, which is the skin current term: $-\mu_0 J_{\|j}^{(\mathrm{skin})}$.

In the above, a part of $\delta A_{\|}$ is solved from an evolution equation and the remainder is solved from the Ampere's law. If $A_{\|}^{(s)}$ carries the dominant part of $\delta A_{\|}$, then $\delta A_{\|}^{(h)}$ will be small, then the skin current will be small, implying that the cancellation error will be small. Then how do we ensure $A_{\|}^{(s)}$ carry the dominant part of $\delta A_{\|}$ over the entire simulation duration? In addition to a careful choice of the evolution equation for $\delta A_{\|}^{(s)}$, we have another leverage that can help $\delta A_{\|}^{(s)}$ to remain dominant: collect the whole $\delta A_{\|}$ into $\delta A_{\|}^{(s)}$ at the end of each time step:

$$\delta A_{\|\mathrm{new}}^{(s)} = \delta A_{\|\mathrm{old}}^{(s)} + \delta A_{\|\mathrm{old}}^{(h)}. \tag{22}$$

Then, to make $\delta A_{\|}$ untouched (so that electromagnetic field remain unchanged), we set $\delta A_{\|}^{(h)}$ to zero:

$$\delta A_{\|\mathrm{new}}^{(h)} = 0. \tag{23}$$

Here "old" and "new" refers to before and after the re-splitting, respectively. The re-splitting keeps the value of $\delta A_{\|}$ untouched and hence does not influence the electromagnetic field. Meanwhile, we need to keep physical $\delta f$ unchanged. The definition of Eq. (18) indicates that, for a given $\delta f$, the re-splitting will make the value of $\delta f^{(h)}$ change as

$$\delta f_{\mathrm{new}}^{(h)} = \delta f_{\mathrm{old}}^{(h)} + \frac{q}{m} \langle v_{\|} \delta A_{\|\mathrm{old}}^{(h)} \rangle \frac{\partial F_0}{\partial \varepsilon}. \tag{24}$$

This is the new initial value for $\delta f^{(h)}$. This step is called "pullback". After this, the physical state of the system remains unchanged.

This scheme makes $\delta A_{\|}^{(h)}$ remain small partially because $\delta A_{\|}^{(s)}$ are set to carry all the value of $\delta A_{\|}$ at the beginning of each time step, and partially because of the careful choice of the evolution equation for $\delta A_{\|}^{(s)}$. If we replace the evolution equation (17) by a more simple one: $\partial \delta A_{\|}^{(s)} / \partial t = 0$, our testing indicates, for the same time step, the original scheme gives correct TAEs while the latter is numerically unstable, indicating a good choice of the evolution equation can allow for larger time-step size. Which choice is good can be problem dependent and is not further explored in this paper.

We also note that the skin current is inverse proportional to the species mass, so this current is dominant by electrons. So the mixed variable pullback scheme may be unnecessary for ion species, but in `TEK`, both ions and electrons are treated with the same scheme (with the FLR effect neglected for electrons).

### A.3 Coordinates

`TEK` uses the filed-aligned coordinates $(\psi, \theta, \alpha)$, where $\psi$ is a magnetic surface label, $\theta$ is an arbitrary poloidal angle, and $\alpha$ is a generalized toroidal angle defined by

$$\alpha \;=\; \phi - \int_0^\theta \hat{q}\, d\,\theta, \tag{25}$$

where $\phi$ is the cylindrical toroidal angle, $\hat{q} = \mathbf{B}_0 \cdot \nabla\phi / \mathbf{B}_0 \cdot \nabla\theta$ is the local safety factor. In terms of $\alpha$, the equilibrium magnetic field is written as $\mathbf{B}_0 = \nabla\Psi \times \nabla\alpha$, which is the Clebsch form ($\Psi$ is the poloidal magnetic flux per rad). So $(\psi, \theta, \alpha)$ coordinates can be called Clebsch coordinates. In $(\psi, \theta, \alpha)$ coordinates, $\partial\mathbf{r}/\partial\theta$ is along the magnetic field line. So these coordinates are often called field-aligned coordinates[29]. Sometimes these coordinates are called "flux tube coordinates".

`TEK` has an option that allows users to choose among 3 kinds of $\theta$: equal-volume, equal-arc-length, and straight-field-line. In `TEK`, the $\theta$ range is chosen to be $[-\pi, +\pi)$, with $\theta = -\pi$ in the high-field-side midplane, and $\theta$ increases along the anti-clockwise direction viewed along $\nabla\phi$.

For the radial coordinate $\psi$, the normalized poloidal magnetic flux is used in `TEK` simulations, but the normalized toroidal magnetic flux $\rho_t$ is used in presenting the results here. For notational ease and to be consistent with other codes that use the field-aligned coordinates, we define $x = \psi$, $y = \alpha$, and $z = \theta$.

### A.4 Initial and boundary conditions

#### A.4.1 Initial conditions

`TEK` uses the $\delta f$ PIC method[30], in which each marker has a weight that is proportional to the value of $\delta f$ at the phase space point of the marker. For each species, the initial value of $\delta f$ is set to be $c_r F_0$, where $c_r$ is a tiny uniform random number among all the markers of that species (typical $c_r$ is chosen to be in $[-10^{-4}, 10^{-4})$). This initial value of $\delta f$ is purely numerical noise, which approaches zero when the marker number approaches infinity.

When doing the initial Monte-Carlo sampling, the phase-space is separated into real space and velocity space, which are sampled independently. For real space sampling, `TEK` adopts uniform distribution in the field-aligned coordinates $(x, y, z)$. For velocity space sampling, `TEK` adopts isotropic Maxwellian of a constant temperature higher than the physical temperature for each species. After particle position and velocity are determined, we calculate the magnetic moment, parallel velocity, and the guiding-center position of each marker. These are then used by the guiding-center pusher as initial conditions.

The initial values of the electromagnetic perturbations, $\delta\Phi$ and $\delta A_{\|}$, do not need to be set because they are solved from the field equations (Poisson's equation and parallel Ampere's law), which do not involve time.

### A.4.2 Boundary conditions for fields

We set $\delta\Phi$ and $\delta A_{\|}$ to be zero at the inner and outer radial boundaries. We impose periodic boundary condition in the toroidal direction, which is physically required if we simulate a full torus. However, for the simulation presented here, only one sixth of the full torus (1/6 toroidal wedge) is used. Then the periodic condition means that the modes captured in the simulations are distanced by six in their toroidal mode numbers (i.e., $n = 0, 6, 12, \ldots$), with all modes between them being artificially eliminated.

As to the boundary condition in the poloidal direction (which is the parallel direction in the Clebsch coordinate system we are using), no artificial condition is imposed. The connecting condition across $\theta = \pm\pi$ is determined by the physical requirement that a field line has $2\pi q$ toroidal angle shift after it finish a poloidal loop. Toroidal interpolation is used to infer the field values at $\theta = +\pi$ from those at $\theta = -\pi$.

### A.4.3 Boundary conditions for particles

The guiding center orbit of each marker is pushed in the $(x, y, z)$ coordinates. Whenever a marker's $z$ exceed the range $[-\pi\colon\pi)$, one $\pm 2\pi$ shift is imposed on $z$ to keep it within $[-\pi\colon\pi)$. A corresponding shift of $y$ by $\mp 2\pi q$ is needed to keep the particle at the same spatial location. The new coordinates of the marker is $(x, y \mp 2\pi q, z \pm 2\pi)$. This process is illustrated in Fig. 26.

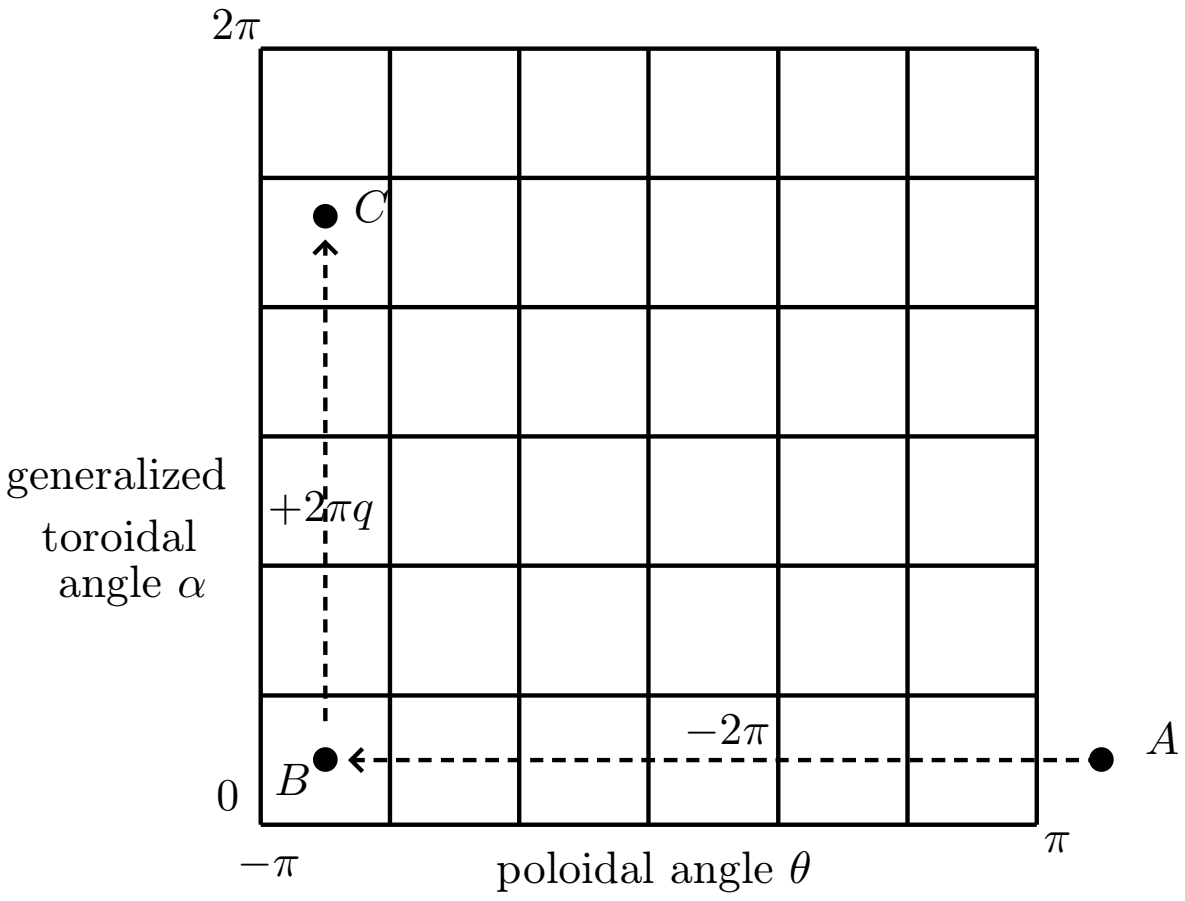

**Figure 26.** To keep the point at the same spatial location when shifting $\theta$ by $-2\pi$, $\alpha$ should be shifted by $+2\pi q$. Here $A$ and $C$ correspond to the same spatial location, but $B$ is at a different location.

We implement two schemes for handling markers' radial boundary condition. The first one is simple: once a marker touches the radial boundary, it is considered as lost and removed from our system. Another is the refill scheme where the weights of those markers touching the radial computational boundary are set to zero and the markers are re-inserted at the mirror (to the midplane) poloidal location.

Whenever a marker's $y$ exceeds the range $[0\colon 2\pi/N_{\rm wedge})$, one $\pm 2\pi/N_{\rm wedge}$ shift is imposed on $y$ to keep it within the range ($N_{\rm wedge}=6$ for the case considered in this paper). This is just the toroidal periodic boundary for markers.

### A.5 Parallelization, gyro-average, pusher, deposition, and field solver

TEK uses one-dimensional MPI (message passing interface) domain decomposition along $z$. Within each $z$ subdomain, markers are further parallelized among MPI processors in the subdomain.

Typical number of gridpoints used in this work is $(N_x, N_y, N_z) = (258, 16, 32)$ with marker number per-cell being 64 for each of the 3 species. Typical MPI processors used are 512, with 16 processors within a single $z$ cell for particle parallelization. Typical time step size is $dt\Omega_{i0} = 4$, where $\Omega_{i0} = B_{0\rm axis} q_i / m_i$, Typical number of time steps is $4 \times 10^4$. Typical wall time for a single-$n$ simulation is about 2 hours.

Marker orbits, weights, and $\delta A_{\parallel}^{(s)}$ are advanced in time by using the second order Rung-Kutta scheme. In each time step, the field equations (Poisson's equation and parallel Ampere's law) are solved twice: one at the half time step and another at the integer time step.

The gyro-ring of each marker is reconstructed in field aligned coordinates $(x, y, z)$ using the 1st order Tailor expansion around the guiding-center location. The gyro-ring is discretized with $N$ points, where $N$ is typically 4. The initial gyro-phase of each marker is set to be random with the purpose of reducing bias (but we did not observe improvement over the nonrandom method). The field values on the $N$ points are obtained by linearly interpolating the field grid values, and are then averaged (gyro-averaging) to give a value that is used to push the marker drift orbit and weight. The gyro-ring is also used when doing the deposition (called scattering in PIC) to get the charge density and current density on the grid. A flat-top particle shape of width equal to the grid spacing is used. This corresponds to linear interpolation when doing the deposition. We use piece-wise constant field reconstruction, then the flat-top shape also means we need to use linear interpolation to get field value at a particle location, as is mentioned above.

In the $y$ direction (toroidal direction), we use the Fourier expansion. When solving the field equations, different toroidal harmonics can be solved independently, which is parallelized by MPI processors in TEK.

We implemented in TEK two different methods of solving the quasi-neutrality condition (Poissons' equation). One is the often used spectrum method (e.g. in GEM code[22, 31]), in which the Fourier spectrum is used to express the polarization density and the double gyro-angle integration is analytically performed, giving rise to the zeroth Bessel's function. Another method is a less mentioned but more straightforward one, in which one directly discretizes the polarization term by numerical integration and using linear interpolation to express it in terms of field values on gridpoints. Simulations performed using the two methods agree with each other, confirming correctness of implementation of the methods in TEK. (The interpolation polarization method has been used in other codes, e.g. GTC code[32, 33, 34].

As to solving the parallel Ampere's law, we use the sine expansion along $x$ for $\delta A_{\parallel}^{(h)}$. The Laplacian operator is approximated by neglecting the parallel derivatives in field-aligned coordinates. More detailed description of the numerical algorithm will be published in another paper.

## B ITG-KBM transition benchmark

Figure 27 compares the results of TKE and GENE for the transition from ion temperature gradient-driven mode (ITG) to kinetic ballooning modes (KBM) in the DIII-D cyclone base case. All parameters are the same as those in T. Gorler's paper[11]. We scan the plasma density and do a series of single-$n$ ($n = 19$) linear simulations. The comparison shows reasonable agreement between the two codes. This demonstrates that TEK can correctly simulate other electromagnetic modes (besides shear Alfven waves discussed in this paper) using the same algorithm.

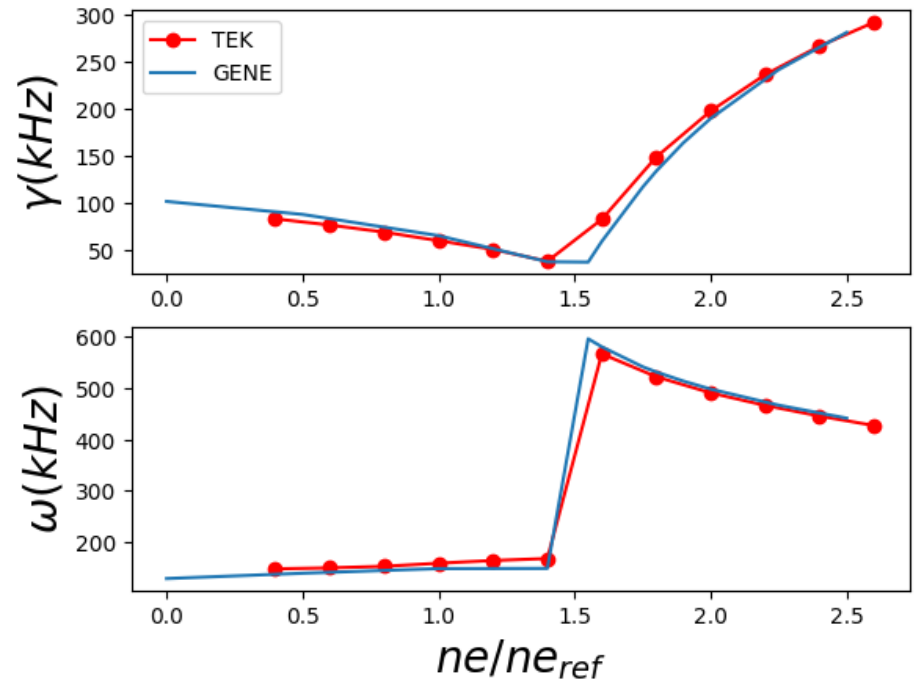

**Figure 27.** Comparison of growth rate and angular frequency between TEK and GENE for the $n = 19$ ITG-KBM transition in the DIII-D cyclone base case. GENE results are from Ref. [11]. The case where $n_e/n_{e,\mathrm{ref}} = 1$ corresponds to the base line case in T. Gorler's paper.